\documentclass[11pt]{article}

\usepackage[latin1]{inputenc}
\usepackage{color}
\usepackage{enumitem}
\usepackage{lscape}
\usepackage{makecell}
\usepackage{lscape}
\usepackage[colorlinks=true, linkcolor=blue, citecolor=blue, urlcolor=blue]{hyperref}
\usepackage{booktabs, multirow}
\usepackage{amssymb,amsmath}
\usepackage{float}
\usepackage{xcolor}
\usepackage[round]{natbib}
\usepackage{authblk}

\usepackage{graphicx}
\usepackage{float}
\usepackage{tabularx}
\usepackage{placeins}
\usepackage{longtable}
\usepackage{arydshln}
\usepackage{tabularray}
\UseTblrLibrary{booktabs}

\usepackage{fancyhdr}
\usepackage{lastpage}

\fancypagestyle{mainstyle}{
	\fancyhf{}
	\fancyfoot[R]{\footnotesize Page \thepage\ of \pageref*{lastmainpage}}

}

\fancypagestyle{suppstyle}{
	\fancyhf{}
	\fancyfoot[R]{\footnotesize Page \thepage\ of \pageref*{LastPage}}

}

\fancypagestyle{first}{
	\fancyhf{}
	\fancyfoot[R]{\footnotesize Page \thepage\ of \pageref*{lastmainpage}}

}

\begin{document}


\title{\textbf{Are continuous models better than areal models for spatial disease mapping with a large number of small areas?}}

\author[1,2]{Salcedo, I.}
\author[1,2]{Trandafir, C.}
\author[1,2]{Goicoa, T.}
\author[3]{Martino, S.}
\author[1,2]{Ugarte, M.D.}

\affil[1]{\small Department of Statistics, Computer Sciences and Mathematics, Public University of Navarre, Spain.}
\affil[2]{\small Institute for Advanced Materials and Mathematics, InaMat$^2$, Public University of Navarre, Spain.}
\affil[3]{\small Department of Mathematical Sciences, Norwegian University of Science and Technology (NTNU).}

\date{}

\makeatletter \pdfbookmark[0]{\@title}{title} \makeatother

\maketitle

\begin{abstract}
\noindent Classical methods for analysing areal mortality data typically rely on Poisson mixed models, which account for spatial and unstructured variability through prior distributions such as the BYM2 prior. 
This paper investigates alternative modeling approaches for settings involving a large number of small areas. Specifically, we explore the Stochastic Partial Differential Equation (SPDE) approach for continuous risk surfaces, examining an existing model from the literature alongside a proposed modification that incorporates an unstructured random effect. 
Finally, we evaluate the performance of both hierarchical mixed Poisson models and SPDE-based frameworks through a comprehensive simulation study across various spatial configurations and aggregation levels.
Furthermore, we apply these approaches to the analysis of lung cancer mortality among men and women across Spanish municipalities during the period 2009--2022.
\end{abstract}


\bigskip

\section{Introduction}\label{sec1}
Disease mapping aims to characterize the geographical distribution of health outcomes, identify spatial patterns of disease risk, and provide evidence to support public health decision-making. In many epidemiological applications, analyses are performed at the small-area level, where the number of observed cases is often low, and the crude standardized mortality or incidence ratios can be highly variable. This problem is particularly pronounced for rare diseases, small municipalities, or analyses stratified by age and sex. This issue has been widely discussed in small-area disease mapping, where sparse event counts and heterogeneous population sizes can lead to unstable crude risk estimates \citep{clayton1987empirical}. In these settings, statistical models are needed to borrow strength between areas and obtain more reliable estimates of the underlying risk surface.

Bayesian hierarchical models for areal data are among the most widely used approaches in disease mapping. These models typically assume that, conditional on the underlying risks, the observed case counts follow a Poisson distribution with an expected-case offset. Spatial random effects are then incorporated to account for residual geographical variation.
Assuming spatial autocorrelation, nearby geographic units are expected to display similar disease risks/rates.  
The Besag-York-Mollié (BYM) model, also named as the convolution model, combines a spatially structured component, specified through an intrinsic conditional autoregressive prior \citep{besag1974}, with an unstructured Gaussian component \citep{besag1991bayesian}. However, the original BYM formulation suffers from identifiability and scaling issues because the structured and unstructured random effects are not separately identifiable. In this paper, we employ the BYM2 model which improves the interpretability and scaling of the model \citep{riebler2016intuitive}.

Despite their usefulness, areal models have some limitations. Spatial dependence is defined through a neighbourhood graph, most often based on shared borders, and the resulting estimates may depend on the chosen adjacency criterion. In addition, areal models are tied to a particular spatial partition. When the same underlying phenomenon is analysed at different geographical resolutions, different risk estimates and conclusions may be obtained, a problem closely related to the modifiable areal unit problem \citep{openshaw1984modifiable}. These limitations are particularly important in studies with many small and irregular areas, such as municipalities, where local instability, spatial aggregation, and computational burden all become relevant. Continuous spatial models provide an alternative way to analyse geographical variation. Instead of defining risk only over a fixed set of regions, they assume an underlying spatial process defined over a continuous domain, which can be modeled as a Gaussian Random Field (GRF). A GRF can be represented in different ways. In this study, we use the stochastic partial differential equation (SPDE) approach as a computationally convenient representation of a GRF with Matérn covariance. The SPDE approach yields a Gaussian Markov Random Field (GMRF) representation on a triangulated mesh, allowing computationally efficient Bayesian inference through sparse matrices \citep{lindgren2011explicit}.

The use of continuous-domain models in disease mapping is not new.  For example, \cite{konstantinoudis2020discrete} used Log-Gaussian Cox processes as precise geocodes were available. However, due to confidentiality issues, individual geocodes are frequently unavailable, and data are instead provided as small-area aggregates. In this paper, we aim to investigate the use of the SPDE approach as an alternative to the BYM2 model, to estimate the effect of both model formulations and spatial support on the estimation of mortality risks in very small areas (Spanish municipalities). The study addresses two related questions. First, we examine whether an areal BYM2 model and continuous-domain SPDE models provide comparable municipal risk estimates. Second, we assess how much spatial information is retained when the data is aggregated into a smaller number of areas. Then, we fit SPDE models at this coarser resolution and use the resulting risk surface to obtain municipal predictions.   
To compare the different modeling approaches, we conduct a comprehensive simulation study  considering different levels of excess risk and expected counts, while separately examining the ability of the aggregated SPDE models to recover the underlying spatial structure at the municipal level. Similar simulation-based studies have been used in disease mapping to evaluate the estimation of the risk surface and the identification of high-risk areas \citep{ugarte2009empirical,richardson2004interpreting}. Then, we analyse male and female lung cancer mortality data for the period 2009--2022, stratified by age group. The main analysis focuses on the 75 year-old or older age group, while results for the younger age groups are provided in the Supplementary Material. 

The paper is organized as follows. Section \ref{section_methods} describes the study design and presents the BYM2 and SPDE models, along with the prior specifications. Section \ref{section_simulation} presents the simulation study designed to evaluate the performance of the models in different spatial risk configurations. In Section \ref{section_results} the different modeling approaches are applied to male and female lung cancer mortality data in Spanish municipalities. Finally, we conclude the article with a discussion of the main findings in Section \ref{section_discussion}.

\section{Methods}\label{section_methods}

\subsection{Study design}\label{subsec1}
The study is based on lung cancer mortality data in Spanish municipalities for the period 2009-2022. For each municipality, the dataset contains the observed number of deaths and the corresponding expected number of deaths obtained from reference mortality rates (in our case the rates of Spain).  The real data analysis presented in Section \ref{section_results} focuses on males and females aged 75 years or older, with separate models fitted for each sex. The original spatial resolution is the municipality, resulting in 7906 areas. Before model fitting, two municipalities with extremely small spatial extent were merged with their nearest neighbouring municipalities. Maleján, located in the province of Zaragoza in Aragón, was merged with Borja, whereas Emperador, located in the province of Valencia in Valencian Region, was merged with Albalat dels Sorells. Their respective areas were $0.031\,\text{km}^2$ and $0.023\,\text{km}^2$. This step was performed to avoid numerical instability in the construction of the spatial mesh and to ensure reliable estimation of the continuous spatial field. The epidemiological counts and geometries of the merged municipalities were aggregated accordingly.

To assess the impact of spatial resolution on model performance, an aggregated partition of 300 areas was constructed similar to \cite{retegui2025}.The aggregation was performed within provinces using a k-means clustering algorithm on the centroid coordinates of the municipalities. This procedure was designed to obtain spatially compact groups while preserving the broad geographical structure of the study region.
For each aggregated area, the municipal geometries were dissolved and the observed and expected counts were summed. We use the resulting partition  to fit the aggregated SPDE models. The predictions from these models were then projected back to the municipal scale, allowing comparison with models fitted directly at that scale.

\subsection{General disease mapping model}\label{subsec2}
Let the study region be divided into $n$ small areas labeled $i = 1, \dots n$. For each area $i$, let $O_i$ and $E_i$ denote the observed and expected number of deaths, respectively. Conditional on the underlying relative risk $r_i$, the number of cases observed in each area $i$ are assumed to follow a Poisson distribution:
\[
O_i \mid r_i \sim \mathrm{Poisson}(E_i r_i),
\qquad i=1,\ldots,n.
\]
The log-relative risk is modeled as
\begin{equation}
	\log(r_i)=\beta_0+\xi_i,
	\qquad i=1,\ldots,n,
	\label{eq:bym2_predictor}
\end{equation}
where $\beta_0$ is the overall log-relative risk and $\xi_i$ is a latent spatial effect accounting for the residual spatial variation.

We consider five spatial modeling strategies to investigate the influence of both the spatial model and the support on which the likelihood is defined. The BYM2 model represents spatial dependence through the municipal neighbourhood structure, whereas the SPDE models assume an underlying continuous risk field.
As described in Section~\ref{subsub_iid}, we consider SPDE model formulations both without and with an additional unstructured random effect. These are denoted as SPDE7904 and SPDE7904 (iid), respectively, reflecting the municipalities retained after the cartographic preprocessing.
Because the BYM2 and municipal SPDE models are fitted to the same counts across all 7904 municipalities, they form the basis of our primary comparison. In addition, we fit SPDE models to 300 aggregated areas, denoted as SPDE300 and SPDE300 (iid), to examine whether models based on coarser observational support can recover the main spatial risk pattern when its predictions are projected back to the municipal level, all while substantially reducing computational cost.

\subsection{Areal model: BYM2}\label{subsec3}
This model represents spatial variation through a combination of spatially structured and unstructured components. To define the structured component, let \(\mathbf{W}=(w_{ij})\) be the adjacency matrix associated with the areal partition, where \(w_{ij}=1\) if areal units \(i\) and \(j\) are neighbours and \(w_{ij}=0\) otherwise. Let \(\mathbf{D}=\operatorname{diag}(w_{i+})\) be the diagonal matrix with entries \(w_{i+}=\sum_j w_{ij}\), representing the number of neighbours of area \(i\). The intrinsic conditional autoregressive (ICAR) precision matrix is then defined as
$\mathbf{Q}=\mathbf{D}-\mathbf{W}.$

The classical BYM model specifies the log-relative risk as
\[
\log(r_i)=\beta_0+u_i+v_i,
\qquad i=1,\ldots,n,
\]
where \(\beta_0\) is an intercept representing the overall level of risk, \(u_i\) is a spatially structured random effect, and \(v_i\) is an unstructured heterogeneity term. The structured component follows an intrinsic autoregressive prior,
\[
u_i \mid u_{-i},\tau_u
\sim
\mathcal{N}
\left(
\frac{\sum_{j=1}^{n}w_{ij}u_j}
{\sum_{j=1}^{n}w_{ij}},
\frac{1}
{\tau_u\sum_{j=1}^{n}w_{ij}}
\right),
\]
whereas the unstructured component is defined as i.i.d. Gaussian noise, $v_i\sim\mathcal{N}(0,\tau_v^{-1}).$

Here, \(\tau_u\) and \(\tau_v\) denote the precision parameters associated with the structured and unstructured effects, respectively. Under the ICAR prior, the spatial effect of each area is shrunk towards the average of its neighbour, whereas \(v_i\) accounts for residual extra-Poisson variability that is not spatially structured. The ICAR prior is improper because its precision matrix \(\mathbf{Q}\) is singular. Consequently, a sum-to-zero constraint, is usually required \citep{besag1995conditional,goicoa2018spatio}. The original BYM formulation also has limitations in interpreting its variance components. In particular, the data provide information primarily on the general variability of \(u_i+v_i\), rather than on the structured and unstructured components separately. Consequently, the variances associated with \(u_i\) and \(v_i\) can be difficult to identify without relying on previous assumptions \citep{macnab2014identification}. Moreover, the marginal scale of the ICAR component depends on the neighbourhood graph, making its precision parameter difficult to interpret and compare between different spatial partitions \citep{sorbye2014scaling}. To address these limitations, \cite{riebler2016intuitive} adopted a scaled reparameterization scheme rooted in the formulation by \cite{dean2001detecting}, known as the BYM2 model. The structured and unstructured components are represented through a single combined effect \(\xi_i\) \eqref{eq:bym2_predictor}.

Let \(\boldsymbol{\xi}=(\xi_1,\ldots,\xi_n)^\top\), and let \(\mathbf{Q}_{*}\) denote the scaled ICAR precision matrix. The scaling is chosen so that the geometric mean of the marginal variances of the structured component is equal to one. The combined effect is defined as

\begin{equation}
\boldsymbol{\xi}\mid\tau_{\xi}\,,\lambda
\sim
\mathcal{N}
\left(
\mathbf{0},
\tau_{\xi}^{-1}\
\left[
(1-\lambda)\mathbf{I}_n
+
\lambda\mathbf{Q}_{*}^{-}
\right]
\right),
\label{eq:bym2}
\end{equation}

\noindent where \(\mathbf{Q}_{*}^{-}\) denotes a generalized inverse of the scaled ICAR precision matrix and $\tau_{\xi}$ is a precision parameter. The overall marginal variance contribution of the combined effect is controlled by \(1/\tau_{\xi}=\sigma_\xi^2\) , while
\(\lambda\in[0,1]\) determines how this variability is distributed between the unstructured and spatially structured components. When \(\lambda=0\), the residual variation is completely unstructured, and the covariance matrix is reduced to
\(\tau_\xi^{-1}\mathbf{I}_n\).
When \(\lambda=1\), the model contains only the scaled ICAR component. Intermediate values of \(\lambda\) combine both sources of residual variability. Scaling the ICAR component makes the interpretation of the marginal variance \(\sigma_\xi^2\) 
less dependent on the particular neighbourhood graph.

\subsection{Continuous spatial models based on SPDE}\label{subsec4}

The areal models described above define spatial dependence through a fixed set of administrative units. An alternative is to assume that the relative risk is generated by an underlying continuous spatial process defined over the whole domain \(\mathcal{D}\). Let \(\mathbf{s}\in\mathcal{D}\) denote a spatial location. We consider two SPDE formulations that differ in whether an area-specific unstructured random effect is included. 

\subsubsection{SPDE model} \label{subsub241}

The point-level log-relative risk is written as
\begin{equation}
\log\{r(\mathbf{s})\}
=
\beta_0 + Z(\mathbf{s}),
\label{eq:continuous_logrisk}
\end{equation}
where \(Z(\mathbf{s})\) is a zero-mean GRF representing the spatially structured component.

In classical geostatistics, Gaussian fields  (GF) are specified through a covariance function. For a finite set of locations \(\mathbf{s}_1,\ldots,\mathbf{s}_n\), the vector $\mathbf{Z}
=
\left(
Z(\mathbf{s}_1),\ldots,Z(\mathbf{s}_n)
\right)^\top$
is assumed to follow a multivariate Gaussian distribution with mean zero and covariance matrix \(\boldsymbol{\Sigma}\), whose elements are
$\Sigma_{ij}
=
\operatorname{Cov}\{Z(\mathbf{s}_i),Z(\mathbf{s}_j)\}.$ This formulation is useful because the covariance function describes how spatial dependence decreases with distance, and its parameters have a direct interpretation. In this work, the covariance is assumed to belong to the Matérn family, one of the most common choices for spatial GF \citep{gneiting2006geostatistical,bakka2018spatial}. The formula of the Matérn covariance function is
\begin{equation}
\operatorname{Cov}\{Z(\mathbf{s}_i),Z(\mathbf{s}_j)\}
=
\frac{\sigma^2}{2^{\nu-1}\Gamma(\nu)}
\left(
\kappa\|\mathbf{s}_i-\mathbf{s}_j\|
\right)^\nu
K_\nu
\left(
\kappa\|\mathbf{s}_i-\mathbf{s}_j\|
\right),
\label{eq:matern_covariance}
\end{equation}
where \(\sigma^2\) is the marginal variance, \(\nu>0\) controls the smoothness of the field, \(\kappa>0\) is a spatial scale parameter, and \(K_\nu(\cdot)\) is the modified Bessel function of order $\nu$.
The parameter \(\kappa\) controls the spatial scale of the Matérn covariance, but it is usually 
more interpretable to express it through a range parameter. Since there is no simple exact relationship, \citet{lindgren2011explicit} use the empirically derived practical range
$
\rho
=
\frac{\sqrt{8\nu}}{\kappa},$
which corresponds approximately to the distance at which the spatial correlation is close to 
0.1, for all values of \(\nu\). Although the covariance-based formulation is intuitive, it becomes computationally intractable when the number of locations is large. The covariance matrix \(\boldsymbol{\Sigma}\) is generally dense because each location is correlated with every other location. As a result, operations such as matrix factorisation and inversion become expensive. GMRF provides a computationally attractive alternative. A GMRF is defined through its precision matrix,
$\mathbf{Q}
=
\boldsymbol{\Sigma}^{-1}$,
and the Markov property induces conditional independence that appears as zeros in \(\mathbf{Q}\). This sparsity allows much faster computations.

The SPDE approach provides a way to link the covariance formulation of GRFs to the computational advantages of GMRFs. The idea is to represent a Matérn GF through an SPDE. Specifically, a field \(Z(\mathbf{s})\) with Matérn covariance can be obtained as the solution of
\[
(\kappa^2-\Delta)^{\alpha/2}
\left\{
\tau Z(\mathbf{s})
\right\}
=
\mathcal{W}(\mathbf{s}),
\qquad
\mathbf{s}\in\mathbb{R}^d,
\]
where \(\Delta\) is the Laplacian operator, \(\mathcal{W}(\mathbf{s})\) denotes Gaussian white noise and \(\tau>0\) is a scale parameter. The parameter \(\alpha\) is related to the smoothness of the Matérn field by
$\alpha
=
\nu + \frac{d}{2}$,
with \(d\) denoting the spatial dimension. In this study, we fixed $\alpha$ = 2, which in two spatial dimensions corresponds to $\nu$ = 1. This representation is useful because the SPDE's parameters are directly related to the usual Matérn parameters in Equation \eqref{eq:matern_covariance}. The parameter \(\kappa\) controls the spatial range, while \(\tau\), together with \(\kappa\), controls the marginal variance. Under this parametrisation,
\[
\sigma^2
=
\frac{\Gamma(\nu)}
{
\Gamma(\alpha)(4\pi)^{d/2}
\kappa^{2\nu}
\tau^2
}.
\]
In practice, the SPDE is solved approximately by constructing a triangulated mesh over the spatial domain and using finite element basis functions. If \(M\) is the number of mesh nodes, the field is approximated as
\[
Z(\mathbf{s})
\approx
\sum_{k=1}^{M}
\psi_k(\mathbf{s}) w_k,
\]
where \(\psi_k(\mathbf{s})\) are piecewise linear basis functions. Each basis function equals 1 at its corresponding mesh node and 0 at the other mesh nodes. The associated basis weights are not independent coefficients, but are defined as a GMRF with a sparse precision matrix that explicitly depends on Matérn parameters \citep{lindgren2015bayesian}. Therefore, the SPDE construction provides a computational representation of the whole continuous spatial process, rather than only the values observed at the data locations.

To connect this continuous latent field with the observed areal data, the point-level risk must be aggregated over each spatial unit. Although the latent risk is defined continuously, the observed data are aggregated over areal units. Therefore, the point-level risk
$
r(\mathbf{s})=\exp\{\beta_0+Z(\mathbf{s})\}
$
must be linked to an areal risk. For an area \(A_i\), this can be done by integrating the continuous risk surface over the area,
$
\int_{A_i}
\exp\{\beta_0+Z(\mathbf{s})\}\,d\mathbf{s}.
$
Since this integral has no analytical solution, it is approximated numerically using integration points \(\mathbf{s}_{ij}\), \(j=1,\ldots,m_i\), associated with \(A_i\), where \(m_i\) is the number of integration points in that area. Let \(a_{ij}\) denote the corresponding integration weights. The latent field $Z(\mathbf{s}_{ij})$ is evaluated through the SPDE finite-element representation, so mesh nodes belonging to triangles that intersect $A_i$ may contribute even if the nodes themselves are outside the area. Mesh resolution determines the spatial scale at which the latent field can vary (See Figure~\ref{fig:spde_meshes}). A finer mesh, with smaller triangles, allows the field to adopt more local variation, whereas a coarser resolution imposes a smoother representation and reduces computational cost. This is particularly relevant for small areas, since several municipalities may be represented within the same or neighbouring mesh triangles. In such cases, the areal integral can still be evaluated even if the municipality does not contain a mesh node, but local contrasts may be attenuated by the limited spatial resolution. The weighted average approximates the areal relative risk
\[
r_i
\approx
\frac{
\sum_{j=1}^{m_i}
a_{ij}
\exp\{\beta_0+Z(\mathbf{s}_{ij})\}
}{
\sum_{j=1}^{m_i}
a_{ij}
},
\]

and the log risk is modelled

\begin{equation}
\log(r_i)
\approx
\log
\left[
\frac{
	\sum_{j=1}^{m_i}
	a_{ij}
	\exp\{\beta_0+Z(\mathbf{s}_{ij})\}
}{
	\sum_{j=1}^{m_i}
	a_{ij}
}
\right].
\label{eq:areal_logrisk_rescaled}
\end{equation}
This predictor \eqref{eq:areal_logrisk_rescaled} is nonlinear because the logarithm is applied after averaging the exponentiated latent field. As a result, it cannot be fitted directly within the standard INLA framework, which assumes a predictor that is linear in the latent components. We therefore use the \texttt{inlabru} framework, which accommodates nonlinear predictors through iterative linearisation \citep{lindgren2024inlabru}. This formulation is useful for the disaggregation setting below, since the same continuous latent field can be linked to different areal supports through the corresponding integration weights.

\begin{center}

\includegraphics[
width=0.65\textwidth
]{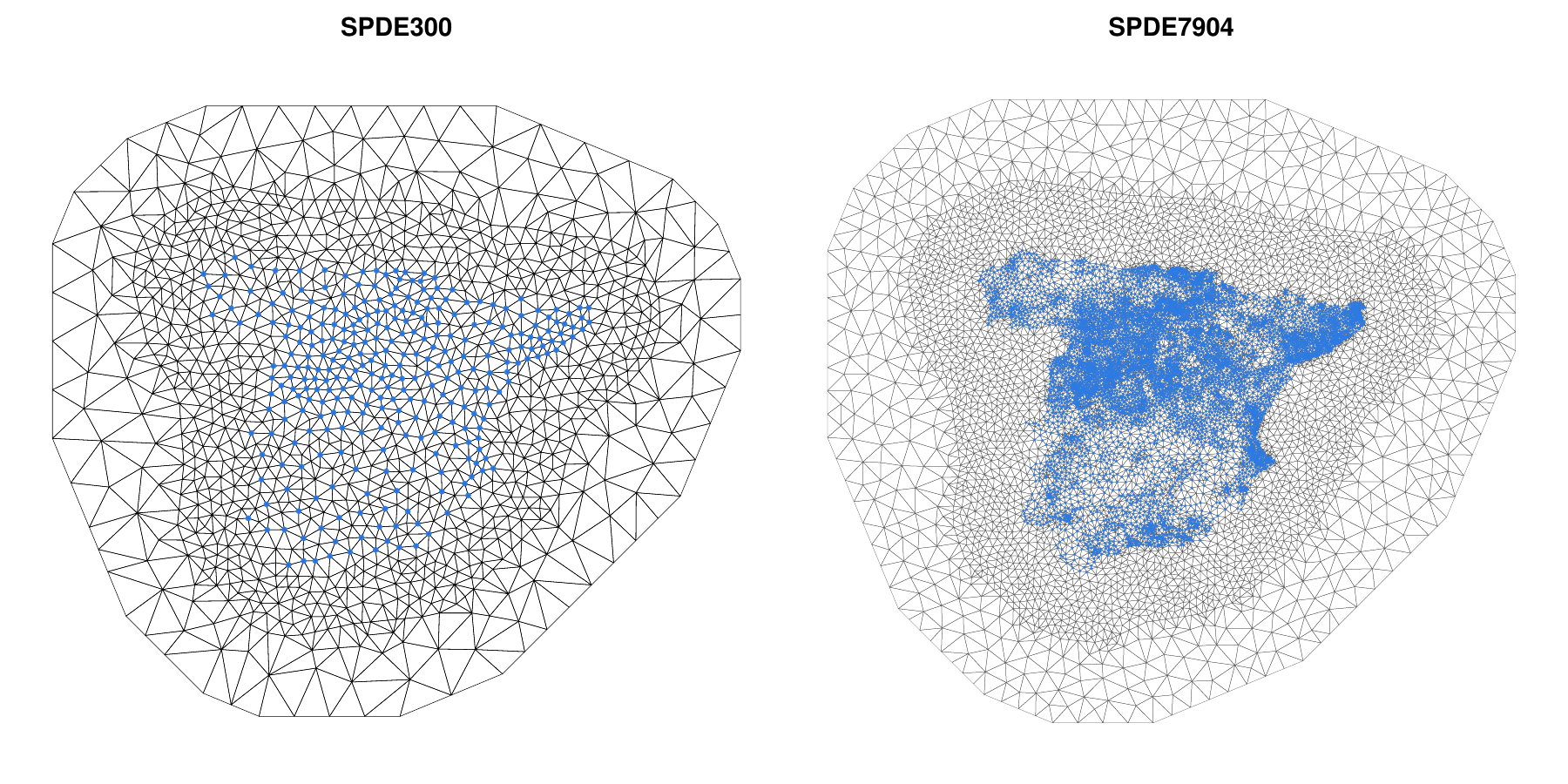}

\vspace{0.5em}

\refstepcounter{figure}
\label{fig:spde_meshes}

\begin{minipage}{\textwidth}
	\small
	\noindent
	\textbf{Figure \thefigure:} Triangulated meshes used for the SPDE models. The left panel shows the mesh used for the SPDE300 models, constructed from the centroids of the aggregated areas, whereas the right panel shows the finer mesh used for the SPDE7904 models, constructed from the municipal centroids. Blue points represent the corresponding centroids of the areas used to construct each mesh. For each spatial support, the same mesh is used for the models with and without iid component.
\end{minipage}

\end{center}

\subsubsection{SPDE model with an unstructured random effect} \label{subsub_iid}
The SPDE formulation described in the previous section represents variation in the relative risk through the continuous field $Z(\mathbf{s})$. This component captures spatial dependence by allowing the risk at nearby locations to be correlated. However, variability between areas may also arise from factors that do not exhibit a spatially structured pattern and cannot be fully represented by the continuous field. To account for this additional source of heterogeneity, we extend the previous formulation by including an independent area-specific random effect. The model therefore combines the continuous spatial field, which accounts for spatially structured variation, with an unstructured component that accounts for area-specific heterogeneity.

The point-level log-relative risk is then defined as
\[
\log\{r(\mathbf{s})\}
=
\beta_0 + Z(\mathbf{s}) + u_i,
\qquad \mathbf{s}\in A_i,
\label{eq:continuous_logrisk_ui}
\]
where \(u_i\) is an independent area-specific random effect. The definition of \(A_i\) depends on the observational support of the model: it corresponds to the \(i\)th municipality for SPDE7904 and to the \(i\)th aggregated area for SPDE300. Thus, while \(Z(\mathbf{s})\) varies continuously over the spatial domain \(\mathcal{D}\), \(u_i\) is common for all locations \(\mathbf{s}\in A_i\). Recall that, hereafter, we denote the models including this unstructured random effect as SPDE7904 (iid) and SPDE300 (iid), respectively.

Following the same areal integration procedure described in Section~\ref{subsub241}, the point-level risk is aggregated over each area $A_i$.
The corresponding areal relative risk is approximated as 
\[
r_i
\approx
\frac{
\sum_{j=1}^{m_i}
a_{ij}
\exp\{\beta_0+Z(\mathbf{s}_{ij})\}
}{
\sum_{j=1}^{m_i}
a_{ij}
}\exp\{u_i\},
\]
and therefore
\[
\log(r_i)
\approx
\log
\left[
\frac{
\sum_{j=1}^{m_i}
a_{ij}
\exp\{\beta_0+Z(\mathbf{s}_{ij})\}
}{
\sum_{j=1}^{m_i}
a_{ij}
}
\right] + u_i.
\label{eq:areal_logrisk_rescaled_ui}
\]

\subsection{Priors specification}\label{subsec5}

All models were fitted within a Bayesian framework. In this study, we use penalized complexity (PC) priors, following \cite{simpson2017penalising}, as they provide a convenient way to control model complexity while maintaining a clear interpretation. For SPDE models, this prior framework is particularly useful because the range and marginal standard deviation have a direct spatial interpretation \citep{fuglstad2019constructing}. 

For the BYM2 model, we adopt the default INLA prior specification. In particular, for the precision parameter  $\tau_{\xi}$ we assign a PC prior defined through
$P(1/\sqrt{\tau_{\xi}} > 1) = 0.01,$
where  $1/\sqrt{\tau_{\xi}}=\sigma_{\xi}$ denotes the marginal standard deviation of the spatial effect.

For the parameter $\lambda$, which determines the relative contribution of structured and unstructured components, we assign a PC prior such that
$P(\lambda < 0.5) = 0.5.$ This is a balanced prior specification, without preference for spatially structured or unstructured variability.

For the SPDE models, we assign PC priors for the Matérn covariance function parameters. In particular, we define the range parameter $\rho$ and the marginal standard deviation $\sigma$ as $P(\rho < \rho_0) = 0.1, \quad P(\sigma > 0.5) = 0.5$, where $\rho_0$ depends on the spatial resolution of the model. For the SPDE model fitted at the municipal level,  we set $\rho_0$ at 100 km, reflecting the larger spatial domain. In contrast, for the aggregated SPDE model with 300 areas, we set $\rho_0$ at 50 km, corresponding to a lower range parameter and consistent with a coarser spatial resolution.

For the unstructured component \(u_i\), we assign an independent Gaussian prior with precision \(\tau_u\), together with a PC prior on its standard deviation 
\(1/\sqrt{\tau_u}\), specified as $P(1/\sqrt{\tau_u} > 1)=0.01$. 
The same prior specification for the unstructured component is used in both SPDE7904 (iid) and SPDE300 (iid).
These values were chosen according to the spatial scale of the study region, allowing enough flexibility while avoiding unrealistically rough spatial fields. All models were implemented in R version 4.5.1 using the \texttt{inlabru} package (version 2.14.0).

\section{Simulation study}\label{section_simulation}

We design a simulation study to evaluate the performance of the five spatial models (BYM2, SPDE300, SPDE300 (iid), SPDE7904, and SPDE7904 (iid)) under controlled risk surfaces. We note that for the two SPDE approaches, we consider both the standard formulation and an extension including an independent and identically distributed (iid) random effect. The additional iid term is included for methodological comparability because BYM2 contains both spatially structured and unstructured components, whereas the standard SPDE formulation represents spatially structured variation through the latent continuous field. The SPDE formulations without the iid component are retained to assess how this additional source of unstructured heterogeneity affects the comparison.
Because the true underlying risk is unknown in real applications, the simulation setting provides a controlled framework in which the accuracy of the estimation and the ability to identify high-risk areas can be directly assessed. Simulation studies are widely used in disease mapping to evaluate these properties under predefined spatial configurations \citep{ugarte2009empirical,richardson2004interpreting}.

In all scenarios, we generate \(B=100\) datasets from a Poisson model using the expected number of cases for males 75 years or older, imitating the setting of the real case study. 
To create realistic risk surfaces, we assign an elevated risk to a subset of municipalities while maintaining a baseline relative risk of 1 for the remainder.
Specifically, we examine an elevated relative risk of 1.2 in detail in the main paper and, to save space, move the results corresponding to elevated relative risks of 1.5 and 2 to the supplementary material. We evaluate three distinct spatial configurations in Scenarios 1 to 3:

\begin{itemize}
	
	\item Scenario 1: Comprises 500 high-risk municipalities randomly distributed throughout the study region. This represents an unstructured spatial configuration in which elevated risks are isolated rather than clustered.
	
	\item Scenario 2: Features a single contiguous high-risk cluster that accounts for approximately 4\% of the total expected cases.
	
	\item Scenario 3: Contains five non-overlapping clusters located in different parts of the study region, jointly representing approximately 20\% of the population exposed to elevated risk. 
\end{itemize}

\noindent The spatial configuration of the three scenarios is displayed in the first column of Figure \ref{fig:simulation_rr_surfaces_iid}.

We evaluate model performance using relative root mean squared error (RRMSE),  mean absolute relative bias (MARB), empirical coverage of the 95\% credible intervals, as well as sensitivity and specificity for detecting high risk areas.  If we denote \(RR_i^{true}\) the true relative risk in the area \(i\) and   \(\widehat{RR}_{ib}\)  the posterior mean estimate in simulation \(b\), \(b=1,\ldots,B\), the relative error is defined as
$ 
RE_{ib}
=
100\,
\frac{\widehat{RR}_{ib}-RR_i^{true}}{RR_i^{true}}.
$ The area-specific relative RMSE is
$
RRMSE_i =
\sqrt{
	\frac{1}{B}\sum_{b=1}^{B} RE_{ib}^2
},
$ 
and the global RRMSE is obtained by averaging \(RRMSE_i\) in all areas. For each area, the relative error is averaged across simulations, and MARB is then computed as the mean of the absolute area-specific average relative errors. Specifically, 
$ 
MARB
=
\frac{1}{n}\sum_{i=1}^{n}
\left|
\frac{1}{B}\sum_{b=1}^{B}RE_{ib}
\right|.
$ 
Coverage is the proportion of simulations in which the 95\% credible interval contains the true risk. For each simulation, true positives (\(TP\)) refer to truly high-risk areas that were correctly classified as high-risk by the model. False positives (\(FP\)) are municipalities with baseline risk that were incorrectly classified as high risk. True negatives (\(TN\)) are municipalities with baseline risk that were not classified as high-risk areas and false negatives (\(FN\)) are truly high-risk municipalities that the model did not detect. Then sensitivity and specificity are computed as

\begin{center}
	\makebox[\linewidth][c]{%
		$\displaystyle
		\mathrm{Sensitivity}=\frac{TP}{TP+FN},
		\qquad 
		\mathrm{Specificity}=\frac{TN}{TN+FP}.
		$
	}
\end{center}

\subsection{Simulation results}\label{subsec:simulation_results}

Table~\ref{tab:simulation_rr12} displays the simulation results obtained using the original expected counts and a true relative risk of 1.2 for high risk areas. Overall, model performance depends strongly on the spatial configuration of the high-risk municipalities.

Scenario~1, in which high-risk municipalities are randomly scattered, BYM2 and SPDE7904 (iid) provide very similar RRMSE and MARB, with coverage close to 0.95, whereas SPDE300 models shows slightly larger estimation errors and lower coverage. Detecting high risk areas is particularly difficult in this scenario: sensitivity is low for all models, although SPDE300 and SPDE300 (iid) reach higher values at the expense of lower specificity and a larger number of false positives. BYM2 and SPDE7904 (iid) show almost identical specificity and a considerably lower number of false positives. 

\begin{table}[!htbp]
	\centering
	\resizebox{\textwidth}{!}{
		\begin{talltblr}[
			caption = {Simulation results for a true relative risk of 1.2 using the original expected counts. RRMSE and MARB are expressed as percentages. Mean FP denotes the mean number of false-positive municipalities across simulations.},
			label = {tab:simulation_rr12}
			]{
				colspec = {c l r r r r r r},
				row{1} = {font=\bfseries},
				column{1} = {halign=c},
				column{2} = {halign=l},
				column{3-8} = {halign=r},
				rows = {font=\fontsize{7.4}{8.6}\selectfont},
				rowsep = 1.5pt
			}
			
			\toprule
			Scenario & Model & RRMSE (\%) & MARB (\%) & 95\% Coverage & Sensitivity & Specificity & Mean FP \\
			\midrule
			
			1 & BYM2     & 2.55 & 2.25 & 0.937 & 0.041 & 0.994 & 42.0 \\
			\cline[dashed]{2-8}
			1 & SPDE300  & 3.37 & 2.51 & 0.921 & 0.169 & 0.854 & 1077.0 \\
			1 & SPDE300 (iid) & 3.07 & 2.41 & 0.921 & 0.186 & 0.842 & 1171.2 \\
			\cline[dashed]{2-8}
			1 & SPDE7904 & 3.13 & 2.32 & 0.935 & 0.099 & 0.946 & 401.1 \\
			1 & SPDE7904 (iid) & 2.57 & 2.24 & 0.937 & 0.040 & 0.994 & 45.1 \\
			
			\midrule
			
			2 & BYM2     & 1.93 & 1.02 & 0.955 & 0.993 & 0.945 & 408.2 \\
			\cline[dashed]{2-8}
			2 & SPDE300  & 2.14 & 0.90 & 0.949 & 0.969 & 0.916 & 626.03 \\
			2 & SPDE300 (iid)  & 2.15 & 0.92 & 0.948 & 0.973 & 0.917 & 624.1 \\
			\cline[dashed]{2-8}
			2 & SPDE7904 & 2.21 & 0.81 & 0.954 & 0.979 & 0.918 & 610.7 \\
			2 & SPDE7904 (iid) & 2.18 & 0.84 & 0.962 & 0.977 & 0.942 & 435.1 \\
			
			\midrule
			
			3 & BYM2     & 4.10 & 3.22 & 0.910 & 0.863 & 0.873 & 805.7 \\
			\cline[dashed]{2-8}
			3 & SPDE300  & 4.06 & 2.37 & 0.912 & 0.974 & 0.770 & 1455.7 \\
			3 & SPDE300 (iid) & 4.04 & 2.38 & 0.923 & 0.974 & 0.781 & 1387.3 \\
			\cline[dashed]{2-8}
			3 & SPDE7904 & 3.94 & 2.55 & 0.865 & 0.980 & 0.735 & 1676.0 \\
			3 & SPDE7904 (iid) & 3.96 & 2.60 & 0.898 & 0.972 & 0.779 & 1398.9 \\
			
			\bottomrule
			
		\end{talltblr}
	}
\end{table}

The differences between models become smaller when the high-risk municipalities form spatial clusters. In Scenario~2, estimation errors are substantially lower than in Scenario~1 and sensitivity is close to one for all models. BYM2 provides the lowest RRMSE and the highest specificity, whereas SPDE7904 models show the lowest MARB. Both SPDE300 formulations also perform well in terms of estimation accuracy, although their specificity is lower than that of BYM2. Thus, all models are able to identify a single high-risk cluster, with SPDE7904 (iid) showing a performance particularly close to BYM2, while SPDE300 models tends to extend the detected high-risk pattern to a larger number of municipalities.

Scenario~3, characterized by several separated high-risk clusters, shows a similar improvement in sensitivity in comparison to Scenario 1. In terms of estimation accuracy, the SPDE models are particularly competitive in this scenario, with slightly lower RRMSE than BYM2. However, the BYM2 model achieves the best tradeoff between sensitivity and specificity, as the SPDE models produce unacceptably high false-positive rates, between 23\% and 27\% \citep[see, for example][]{richardson2004interpreting}.

Comparing the two SPDE formulations, the effect of including the unstructured component is more noticeable for SPDE7904 than for SPDE300. For SPDE300, both formulations give similar results across the three scenarios. For SPDE7904, the inclusion of the iid effect leads to a clear improvement in Scenario~1, particularly in terms of RRMSE, specificity and number of false positives. In Scenarios~2 and~3, differences in estimation accuracy are smaller, although SPDE7904 (iid) provides higher specificity and fewer false positives, particularly in Scenario~2, where its results are close to those obtained with BYM2.

\begin{figure}[!ht]
	\centering
	\includegraphics[width=\textwidth,
	height=0.55\textheight,
	keepaspectratio]{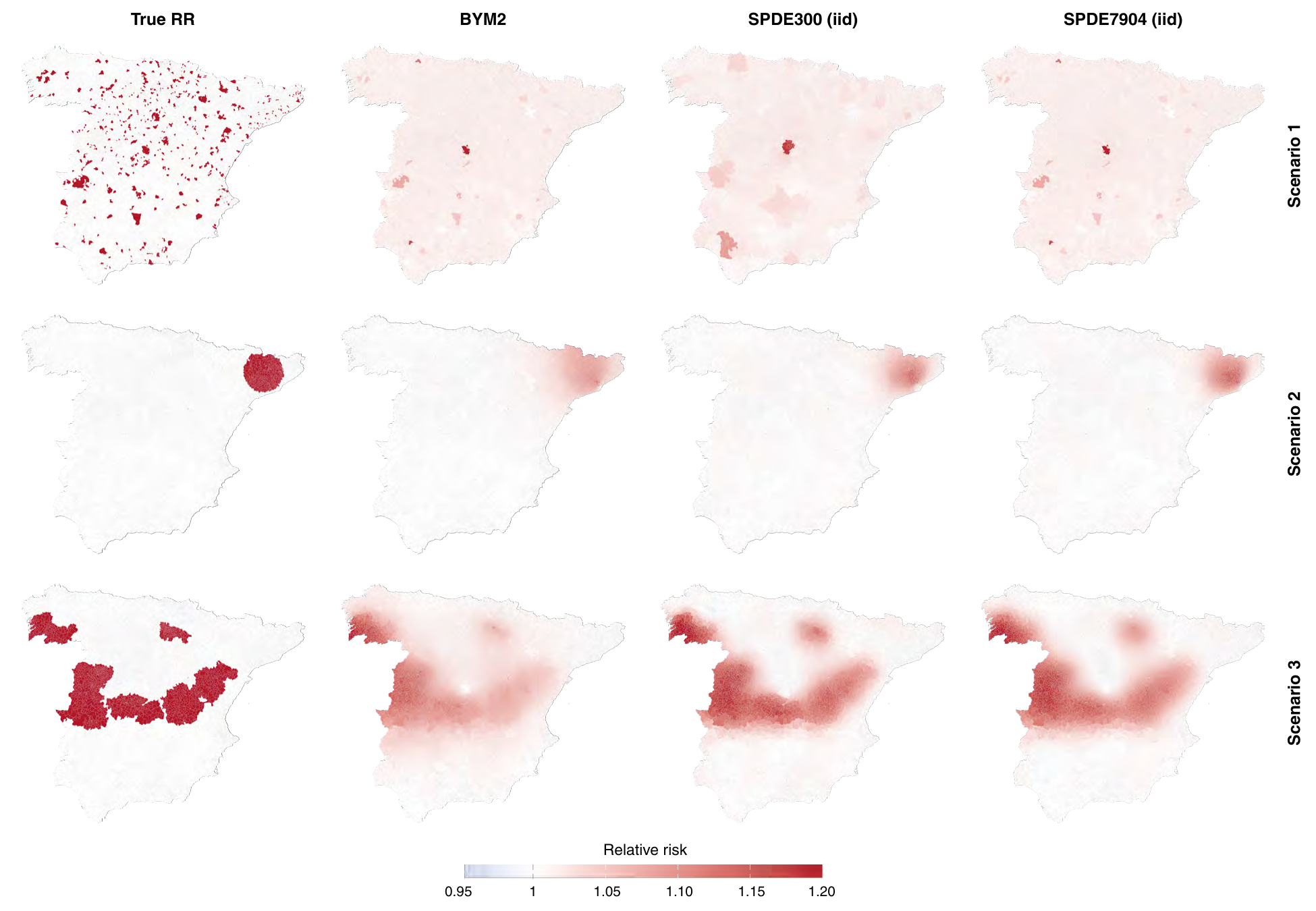}
	\caption{True and mean estimated relative risk surfaces for a true relative risk of 1.2 using the original expected counts. Rows correspond to the three simulation Scenarios. The first column shows the true relative risk, while the remaining columns show the mean municipal relative risk estimated across the 100 simulations.}
	\label{fig:simulation_rr_surfaces_iid}
\end{figure}

Figure~\ref{fig:simulation_rr_surfaces_iid} complements these results by showing the true and mean estimated relative risk surfaces for a true relative risk of 1.2 using the original expected counts. In Scenario~1, BYM2 reproduces some of the isolated high-risk municipalities but with substantial shrinkage towards the baseline, while SPDE7904 (iid) produces a smoother surface in which several local excesses remain visible. SPDE300 (iid) shows the strongest spatial smoothing, with the scattered high-risk municipalities being replaced by broader areas of moderate high risk. In Scenario~2, all three models clearly recover the location of the single high-risk cluster. BYM2 provides a relatively localized estimate around the true cluster, while the SPDE models produce smoother spatial transitions around the cluster boundaries. This smoothing is more pronounced for SPDE300 (iid), whereas SPDE7904 (iid) provides a spatial pattern relatively similar to that obtained with BYM2. In Scenario~3, all models recover the main high-risk regions, although their spatial representation differs. BYM2 identifies the overall clustered structure but shows greater local heterogeneity.
The two SPDE models produce smoother surfaces that visually seem to resemble the shape of the true high-risk regions more closely. SPDE7904 (iid) provides a well-defined representation of the clustered pattern, whereas SPDE300 (iid) shows slightly broader transitions between elevated- and baseline-risk areas.

To complete the simulation study, we also evaluated elevated relative risks of 1.5 and 2, for which the same general patterns were observed. Additional simulations were also performed with expected counts multiplied by factors of 2, 5 and 10,  although the SPDE (iid) models were not considered for these additional simulations to limit computational time (see Tables~\ref{tab:simulation_scenario1},~\ref{tab:simulation_scenario2} and~\ref{tab:simulation_scenario3}  of the Supplementary Material for Scenarios 1, 2, and 3, respectively). Overall, the results are consistent with those obtained using the original expected counts and are therefore not shown here for brevity. Scenario~1 remains the most challenging setting, whereas in Scenarios~2 and~3 sensitivity approaches one as the true relative risk increases and differences in estimation accuracy between models remain relatively small.

In summary, model performance depends on the spatial configuration of the underlying risk.  Overall, the BYM2 model provides lower rates of false-positives (1-specificity),  it remains computationally more efficient than SPDE7904 (as discussed in
Section~\ref{section_results}) and it is competitive in terms of bias and error.

\section{Real data analysis}\label{section_results}

In this section, we analyse lung cancer mortality among males and females aged 75 years or older during the period 2009--2022. A total of 90,153 deaths among males and 20,340 among females were recorded across the 7,904 municipalities during this period. The municipal death counts were highly heterogeneous, with an average number of counts of 11.41 for males and 2.57 for females.

Five models were fitted to the data: BYM2, SPDE7904, SPDE7904 (iid), SPDE300, and SPDE300 (iid). Table~\ref{tab:model_comparison_75} summarizes some model assessment criteria for the 75+ age group, presented separately for males and females. Specifically, the WAIC, the mean log-score and the percentage of high-risk municipalities. High-risk areas were defined using a cut-off value of 0.8 on the posterior exceedance probability, namely $\Pr(RR>1)>0.80$.

\begin{table}
	\centering
	\begin{talltblr}[
		caption = {Model comparison for the 75+ age group by sex.},
		label = {tab:model_comparison_75}
		]{
			colspec = {l l r r r},
			row{1} = {font=\bfseries},
			column{1} = {halign=l},
			column{2} = {halign=l},
			column{3-5} = {halign=r}
		}
		\toprule
		Sex & Model & WAIC & Mean Log-score & \% of high-risk areas \\
		\midrule
		\SetCell[r=5]{m} Males
		& BYM2     & 27348.73 & 1.75 & 9.88  \\
		\cline[dashed]{2-5}
		& SPDE7904 & 27420.16 & 1.74 & 16.21 \\
		& SPDE7904 (iid) & 27352.04 & 1.74 & 13.73 \\
		\cline[dashed]{2-5}
		& SPDE300  & 2473.28  & 4.36 & 19.32 \\
		& SPDE300 (iid) & 2442.92  & 4.34 & 19.56 \\
		\midrule
		\SetCell[r=5]{m} Females
		& BYM2     & 14727.06 & 0.94 & 2.24 \\
		\cline[dashed]{2-5}
		& SPDE7904 & 14789.85 & 0.94 & 5.90 \\
		& SPDE7904 (iid) & 14722.66 & 0.94 & 2.96 \\
		\cline[dashed]{2-5}
		& SPDE300  & 1866.83  & 3.23 & 7.63 \\
		& SPDE300 (iid) & 1845.51  & 3.20 & 7.55 \\
		\bottomrule
	\end{talltblr}
\end{table}

BYM2 and the two SPDE7904 formulations are fitted using the municipal partition, and their model assessment criteria can therefore be compared at the same spatial scale. For both sexes, the inclusion of the iid effect improves the WAIC of SPDE7904 models. For males, the WAIC decreases from 27420.16 for SPDE7904 to 27352.04 for SPDE7904 (iid), bringing it close to the value obtained with BYM2 (27348.73). For females, it decreases from 14789.85 to 14722.66, with SPDE7904 (iid) yielding a slightly lower WAIC than BYM2 (14727.06). The mean log-score is very similar between the two SPDE7904 formulations.

The interpretation of the SPDE300 results differs because these models are fitted using a likelihood defined over 300 aggregated areas rather than over the 7,904 municipalities. Consequently, their WAIC and log-score values are not directly comparable with those of BYM2 and SPDE7904. However, the two SPDE300 formulations are fitted at the same spatial support and can be compared with each other. The inclusion of the iid effect also results in a lower WAIC for both sexes, decreasing from 2473.28 to 2442.92 for males and from 1866.83 to 1845.51 for females.

These results are consistent with the simulation study, where the inclusion of the iid effect was particularly beneficial for the SPDE7904 models, while the differences between the two SPDE300 formulations were smaller. Therefore, we use the SPDE models including the iid effect for the subsequent spatial representation and posterior predictive assessment. The following results focus on BYM2, SPDE7904 (iid) and SPDE300 (iid).

Discrepancies between these models are more pronounced when considering the number of municipalities classified as high risk. Across both sexes, SPDE7904 (iid) identifies a higher percentage of high-risk municipalities than the BYM2 model, a pattern most pronounced among males, whereas SPDE300 (iid) yields the highest overall percentage.

The behaviour of SPDE300 (iid) is related to the spatial support used in the likelihood. By pooling information between municipalities, SPDE300 (iid) reduces the sampling variability associated with small counts in sparsely populated areas. However, this gain in stability comes at the cost of reduced local spatial detail. When the resulting continuous field is projected back to the municipal scale, elevated posterior probabilities extend over larger areas. Consequently, a larger number of municipalities may exceed the fixed posterior probability threshold, even though the model is not fitted directly at the municipal scale.

\begin{figure}[H]
	\centering
	\includegraphics[
	width=\textwidth,
	height=0.50\textheight,
	keepaspectratio
	]{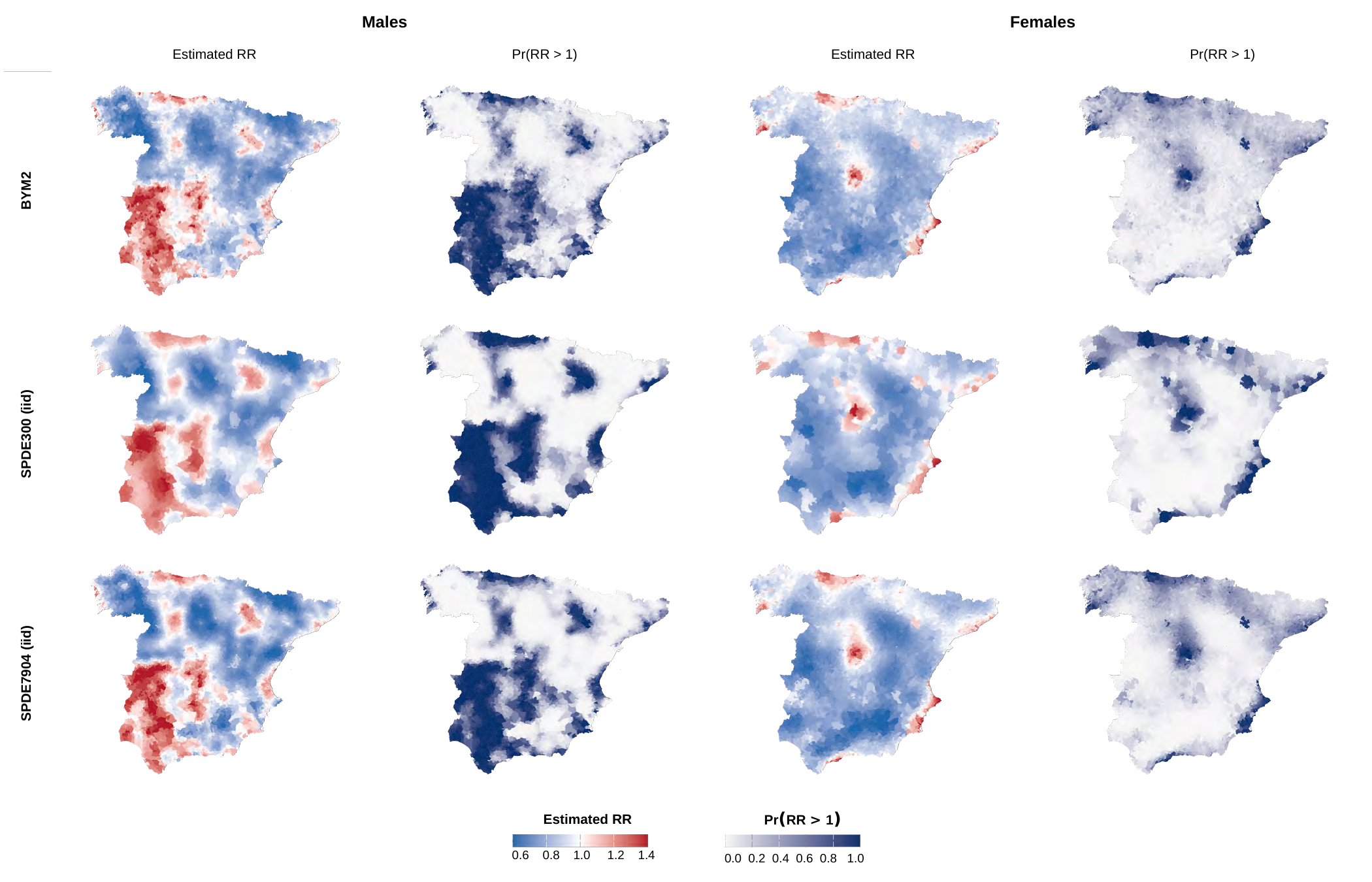}
	\caption{Estimated relative risk of lung cancer mortality and posterior probabilities in males and females aged 75 or older.}
	\label{fig:mapas_hombres}
\end{figure}

Figure~\ref{fig:mapas_hombres} shows the posterior mean relative risks and posterior exceedance probabilities, $\Pr(RR>1)$, for BYM2 (first row), SPDE300 (iid) (second row) and SPDE7904 (iid) (bottom row) for males and females (left and right, respectively). For males, all three models reveal a clear spatial pattern. The highest posterior mean relative risks are concentrated in western and southwestern Spain, particularly in Extremadura and western Andalusia. Elevated risks are also observed in several municipalities in central Spain. BYM2 and SPDE7904 (iid) identify broadly similar high-risk regions, although the SPDE7904 (iid) surface exhibits slightly smoother transitions between neighbouring areas. SPDE300 (iid) produces the smoothest risk surface, reflecting the coarser spatial resolution at which the model is fitted. For females, the spatial pattern is less pronounced than for males. Most municipalities exhibit a low-to-moderate relative risk, while areas with higher posterior means are more localized. The most prominent elevated risks appear in northern Spain, particularly in the Basque Country, Asturias and Cantabria, as well as in parts of central Spain, especially around Madrid, and on the coast of the Valencian Region, Catalonia and Andalusia.

The posterior probability maps complement the relative risk estimates by showing where the evidence for excess mortality is strongest. In males, high values of $\Pr(RR>1)$ are concentrated mainly in western and southwestern areas, with additional areas of elevated probability in central Spain. Among females, the posterior probabilities also exhibit a clear spatial structure, but areas with strong evidence of excess risk are fewer and more geographically restricted than among males. In particular, elevated probabilities are concentrated mainly around Madrid, in several municipalities along the Valencian, Catalan and Andalusian coast, and in a limited number of areas in northern Spain.

\begin{figure}[H]
	\centering
	\includegraphics[
	width=\textwidth,
	height=0.35\textheight,
	keepaspectratio
	]{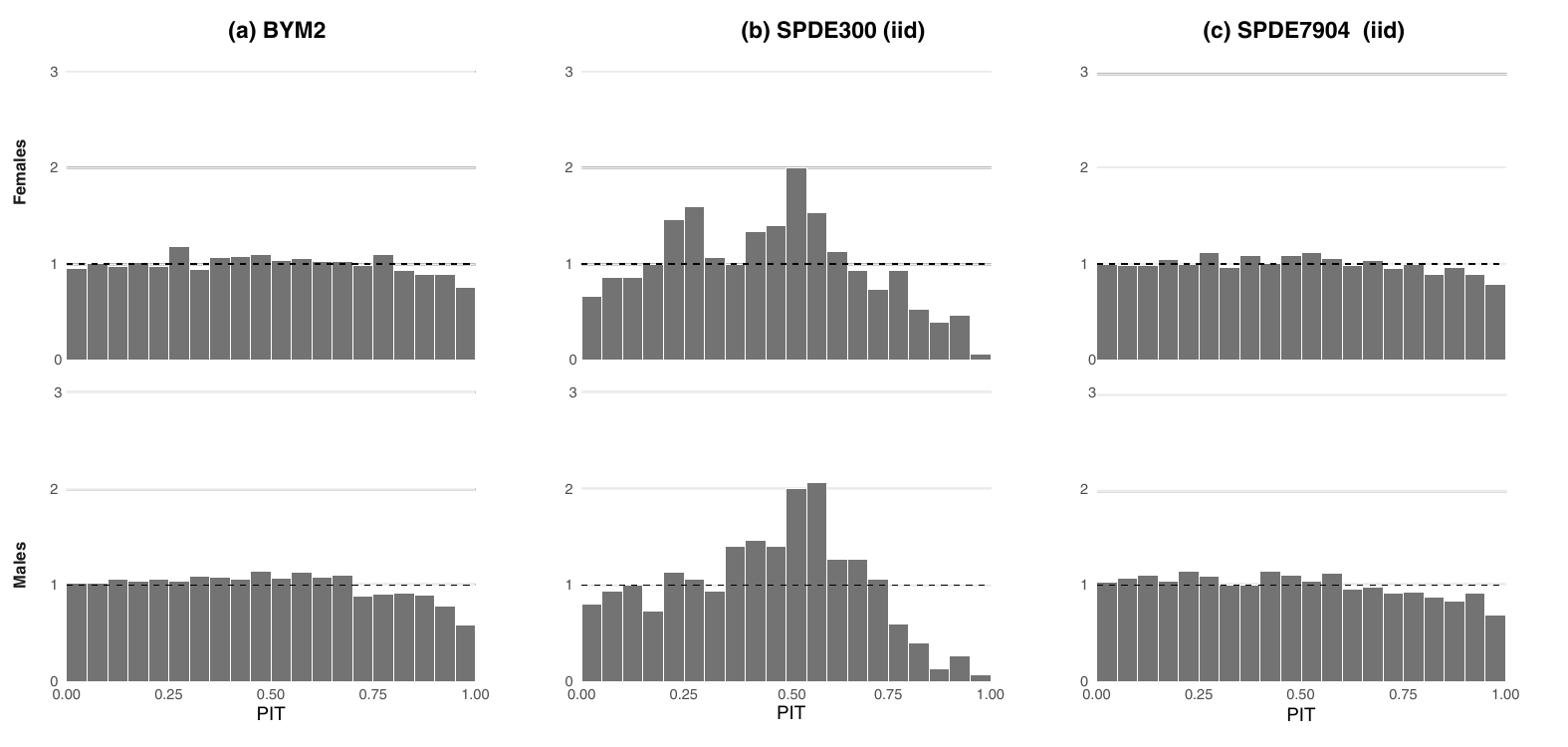}
	\caption{PIT histograms for the 75+ age group for the fitted models, separately by sex.}
	\label{fig:curves_pit_75}
\end{figure}

Figure~\ref{fig:curves_pit_75} displays the probabilistic integral transform (PIT) histograms \citep{dawid1984present} for the age group 75+, which serve as a check on posterior predictive calibration. For each municipality, the PIT compares the observed count with its posterior predictive distribution; for count data, an adjusted or randomized PIT is needed \citep{czado2009predictive}. A well-calibrated model should produce a histogram that is close to a uniform distribution. In the 75+ age group, BYM2 and SPDE7904 (iid) show similar patterns. SPDE300 (iid) behaves differently when evaluated at the municipal level, which is not surprising because it is fitted to aggregated counts and then projected back to municipalities. For this reason, the PIT should be read as a diagnostic of municipal-level predictions, rather than as a direct comparison between models fitted on the same likelihood.

The computational time differs substantially between models. BYM2 takes approximately 4 minutes, the SPDE300 models around 3 minutes, and the SPDE7904 models approximately 1.5 hours for a single analysis on an Intel(R) Xeon(R) Silver 4316 processor with 80 CPUs at 2.30 GHz and 256 GB of RAM.

The Supplementary Material provides additional results for younger age groups. Figures~\ref{fig:supp_maps_45_59} and~\ref{fig:supp_maps_60_74} present the corresponding maps for age groups 45--59 and 60--74.

\FloatBarrier
\section{Discussion and conclusions}\label{section_discussion}
In this paper, we consider two alternative approaches to disease mapping  when the number of small areas increases substantially: the widely used BYM2 model, based on an areal representation of spatial dependence, and continuous-domain SPDE models, with and without an additional unstructured random effect. As a complementary analysis, we assess the effect of spatial aggregation within the SPDE approach on model performance at the small area level. We address these objectives through an extensive simulation study, in which the true underlying risk surface is known, and the analysis of lung cancer mortality among people aged 75 years or older in Spanish municipalities. Our findings highlight the strengths and limitations of areal and continuous-domain representations of spatial dependence for disease mapping. The simulation study shows that the performance of the model  strongly depends on the spatial configuration of the underlying risk surface. When municipalities with elevated risk are spatially isolated, both BYM2 and SPDE7904 (iid) show very similar estimation and detection performance, although the identification of high-risk municipalities remains a challenge for both models, particularly when the relative risk and expected cases are low. As elevated-risk areas form clusters, both approaches achieve high sensitivity and comparable estimation accuracy, with SPDE7904 (iid) showing particularly competitive performance in clustered settings. Overall, the differences between BYM2 and SPDE7904 (iid) are relatively small between simulation scenarios, although BYM2 remains both computationally more efficient and more specific. That is, it provides a lower rate of false positives. Beyond the relative performance of the two approaches, the simulation study also shows that accurate risk estimation and the identification of high-risk municipalities are not always equivalent objectives. In general, a very large bias usually translates into a poor detection of high risk areas (see Table~\ref{tab:simulation_risk_areas_s1} in the Supplementary Material, corresponding to Scenario 1). However, in some simulation settings, municipalities with elevated risk are correctly identified despite noticeable estimation errors, indicating that a moderate bias in risk estimates can still lead to a good detection of high risk areas (see Tables~\ref{tab:simulation_risk_areas_s2} and~\ref{tab:simulation_risk_areas_s3}  of the Supplementary Material, corresponding to Scenarios 2 and 3). This suggests that classification depends more on the ability of a model to distinguish municipalities with elevated risk from the surrounding background areas than on the precise estimation of the magnitude of the relative risk. Consequently, models that show similar classification performance may still differ in estimation accuracy, particularly within high-risk areas. These findings highlight that estimation accuracy and classification performance capture complementary aspects of model behavior and should therefore be interpreted jointly when comparing disease mapping models.

The behavior of SPDE300 models further illustrates the influence of spatial resolution on disease mapping. Unlike SPDE7904 models, which is fitted directly at the municipal level, SPDE300 models are estimated on an aggregated spatial support before predictions are projected back to municipalities. This reduction in spatial resolution produces smoother estimates, although the dominant geographical patterns are largely preserved.  However, its ability to detect truly localized excess risks is reduced, leading to a substantial increase in false positives, a serious issue as false alarms may lead to unwarranted public concern. This is because the posterior probabilities estimated at the level of large areas are propagated to the municipalities within them, limiting the method?s ability to detect truly localized excess risks and potentially resulting in a substantial number of false positives. These findings indicate that the loss of municipality-level accuracy is primarily associated with fitting the likelihood on a coarser spatial support rather than with the SPDE formulation itself. Consequently, SPDE300 models substantially reduced computational cost while preserving the overall geographical structure of the estimated risk surface, making it a useful alternative for exploratory analyses or applications focused on broad spatial patterns, but it is less appropriate when reliable municipality-level inference or the identification of individual high-risk municipalities is required.

The real data analyses largely support the conclusions drawn from the simulation study. BYM2 and SPDE7904 models produce similar general spatial patterns of lung cancer mortality risk, despite relying on different representations of spatial dependence. However, SPDE7904 models identify a larger number of municipalities with high posterior probabilities than BYM2, showing that the two models differ more clearly in the extent of the areas classified as high risk than in the general pattern. In contrast, SPDE300 models produce smoother risk surfaces and more spatially extensive areas with high posterior probabilities, consistent with the behaviour observed in the simulation study and confirming that fitting the likelihood to an aggregated spatial support primarily affects local spatial detail rather than the broader geographical distribution of risk.

The spatial patterns identified in the real data are consistent with previous studies of lung cancer mortality in Spain. Earlier analyses reported high-risk areas among men in the south-west, especially Huelva, Seville and Cádiz, as well as in northern provinces such as Asturias and Cantabria \citep{cayuela2020analisis}. In the present study, both men and women exhibit a marked spatial structure, although the extent and distribution of high-risk areas differ by sex. Among men, high posterior probabilities extend in more municipalities, with a persistent pattern in western and south-western Spain and additional high-risk areas in central Spain among those aged 75 years or older. This distribution is consistent with the earlier and more widespread smoking epidemic among Spanish men and may also reflect the historical distribution of occupational and socioeconomic exposures, particularly in industrial and port areas. In contrast, high-risk municipalities among women are fewer and more geographically restricted, with elevated risks in northern Spain and, among the oldest women, around Madrid and in parts of the Valencian, Catalan and Andalusian coast. This pattern is consistent with previous studies showing that lung cancer mortality among Spanish women has a more localized distribution than that observed in men \citep{retegui2021estimating,cayuela2020analisis}.

The differences observed by sex and age should also be interpreted in the context of the long latency of lung cancer and the different historical patterns of tobacco exposure in Spain. Although lung cancer mortality remains higher among men, recent studies have reported declining mortality rates in men but increasing trends among women. These trends are consistent with the later uptake of smoking among women and with differences in smoking cessation by sex, age, and socioeconomic status \citep{medin2025spatiotemporal,jimenez2011mujeres}. The elevated risks observed around Madrid and along parts of the Valencian coast may also reflect the earlier adoption of smoking among women living in more urbanised areas. In particular, previous cohort studies reported an increase in lung cancer mortality among Spanish women born between 1935 and 1955 \citep{bramajo2024lung}, suggesting that the spatial pattern observed among the oldest women likely reflects smoking behaviours established earlier in life. Although smoking remains the main risk factor for lung cancer, previous studies indicate that population density, socioeconomic conditions, occupational exposures, environmental factors, and mining activity may also contribute to the geographical distribution of lung cancer mortality in Spain \citep{medin2025spatiotemporal}.

In summary, the results indicate that the relative performance of areal- and continuous-domain formulations depends on the characteristics of the underlying spatial risk process. BYM2 combines high accuracy and computational efficiency, while SPDE7904 (iid) achieves comparable performance when elevated-risk municipalities are randomly distributed or form a single spatial cluster. For some spatial configurations, however, SPDE7904 (iid) may still produce substantially more false positives than BYM2. This is an important consideration in disease mapping, as excessive false-positive signals may lead to the identification of areas as being at elevated risk despite insufficient evidence, potentially generating unwarranted public health alerts. Although fitting the SPDE model on an aggregated spatial support substantially reduces computational cost, this is accompanied by a loss of municipality-level detail. Taken together, these findings demonstrate that the relative performance of spatial disease mapping models is determined not only by the modeling framework itself but also by the characteristics of the underlying spatial process and the spatial resolution at which the model is fitted. Consequently, the choice of modeling approach should therefore be  guided by the spatial resolution required for inference, the available computational resources, and the control of false positives.

\section*{Code availability}

The code supporting this study will be made available upon acceptance in a
\href{https://github.com/spatialstatisticsupna/Areal_vs_continuous}{GitHub repository}.
For confidentiality reasons, synthetic data will be provided imitating the real data.

\section*{Acknowledgements}

This work has been supported by PID2024-155382OBI00/MICIU/AEI/10.13039/501100011033 funded by the Ministry of Science, Innovation and Universities of Spain and FEDER (UE).

\pagebreak

\bigskip

\bibliographystyle{apalike}
\bibliography{bibliography}   

@article{riebler2016intuitive,
  title={An intuitive Bayesian spatial model for disease mapping that accounts for scaling},
  author={Riebler, Andrea and S{\o}rbye, Sigrunn H and Simpson, Daniel and Rue, H{\aa}vard},
  journal={Statistical Methods in Medical Research},
  volume={25},
  number={4},
  pages={1145--1165},
  year={2016},
  publisher={SAGE Publications Sage UK: London, England}
}

@article{besag1991bayesian,
  title={Bayesian image restoration, with two applications in spatial statistics},
  author={Besag, Julian and York, Jeremy and Molli{\'e}, Annie},
  journal={Annals of the Institute of Statistical Mathematics},
  volume={43},
  number={1},
  pages={1--20},
  year={1991},
  publisher={Springer}
}

@article{goicoa2018spatio,
  title={In spatio-temporal disease mapping models, identifiability constraints affect {PQL} and {INLA} results},
  author={Goicoa, Tom{\'a}s and Adin, Aritz and Ugarte, Mar{\'\i}a Dolores and Hodges, James S},
  journal={Stochastic Environmental Research and Risk Assessment},
  volume={32},
  number={3},
  pages={749--770},
  year={2018},
  publisher={Springer}
}

@article{besag1995conditional,
  title={On conditional and intrinsic autoregressions},
  author={Besag, Julian and Kooperberg, Charles},
  journal={Biometrika},
  volume={82},
  number={4},
  pages={733--746},
  year={1995},
  publisher={Oxford University Press}
}

@article{lindgren2011explicit,
  title={An explicit link between {Gaussian} fields and {Gaussian} {Markov} random fields: the stochastic partial differential equation approach},
  author={Lindgren, Finn and Rue, H{\aa}vard and Lindstr{\"o}m, Johan},
  journal={Journal of the Royal Statistical Society Series B: Statistical Methodology},
  volume={73},
  number={4},
  pages={423--498},
  year={2011},
  publisher={Oxford University Press}
}

@article{lindgren2024inlabru,
  title={inlabru: software for fitting latent {Gaussian} models with non-linear predictors},
  author={Lindgren, Finn and Bachl, Fabian and Illian, Janine and Suen, Man Ho and Rue, H{\aa}vard and Seaton, Andrew E},
  journal={arXiv preprint arXiv:2407.00791},
  year={2024}
}

@article{simpson2017penalising,
  title={Penalising model component complexity: A principled, practical approach to constructing priors},
  author={Simpson, Daniel and Rue, H{\aa}vard and Riebler, Andrea and Martins, Thiago G and S{\o}rbye, Sigrunn H},
  year={2017}
}

@article{ugarte2009empirical,
  title={Empirical {Bayes} and Fully {Bayes} procedures to detect high-risk areas in disease mapping},
  author={Ugarte, Mar{\'\i}a Dolores and Goicoa, Tom{\'a}s and Militino, Ana F},
  journal={Computational Statistics \& Data Analysis},
  volume={53},
  number={8},
  pages={2938--2949},
  year={2009},
  publisher={Elsevier}
}

@article{dean2001detecting,
  title={Detecting interaction between random region and fixed age effects in disease mapping},
  author={Dean, CB and Ugarte, MD and Militino, AF},
  journal={Biometrics},
  volume={57},
  number={1},
  pages={197--202},
  year={2001},
  publisher={Wiley Online Library}
}

@article{openshaw1984modifiable,
  title={The modifiable areal unit problem},
  author={Openshaw, Stan},
  journal={Concepts and techniques in modern geography},
  year={1984},
  publisher={GeoBooks}
}

@article{retegui2021estimating,
  title={Estimating LOCP cancer mortality rates in small domains in {Spain} using its relationship with lung cancer},
  author={Retegui, Garazi and Etxeberria, Jaione and Ugarte, Mar{\'\i}a Dolores},
  journal={Scientific Reports},
  volume={11},
  number={1},
  pages={22273},
  year={2021},
  publisher={Nature Publishing Group UK London}
}

@article{retegui2025,
  title={On prior smoothing with discrete spatial data in the context of disease mapping},
  author={Retegui, Garazi and Gelfand, Alan and and Etxeberria, Jaione and Ugarte, Mar{\'\i}a Dolores},
  journal={Statistical Methods in Medical Research},
  volume={34},
  number={10},
  pages={2091--2107},
  year={2025},
  publisher={Sage}
}

@article{richardson2004interpreting,
  title={Interpreting posterior relative risk estimates in disease-mapping studies},
  author={Richardson, Sylvia and Thomson, Andrew and Best, Nicky and Elliott, Paul},
  journal={Environmental Health Perspectives},
  volume={112},
  number={9},
  pages={1016},
  year={2004}
}

@article{czado2009predictive,
  title={Predictive model assessment for count data},
  author={Czado, Claudia and Gneiting, Tilmann and Held, Leonhard},
  journal={Biometrics},
  volume={65},
  number={4},
  pages={1254--1261},
  year={2009},
  publisher={Oxford University Press}
}

@article{dawid1984present,
  title={Present position and potential developments: Some personal views statistical theory the prequential approach},
  author={Dawid, A Philip},
  journal={Journal of the Royal Statistical Society: Series A (General)},
  volume={147},
  number={2},
  pages={278--290},
  year={1984},
  publisher={Wiley Online Library}
}

@article{konstantinoudis2020discrete,
  title={Discrete versus continuous domain models for disease mapping},
  author={Konstantinoudis, Garyfallos and Schuhmacher, Dominic and Rue, H{\aa}vard and Spycher, Ben D},
  journal={Spatial and Spatio-temporal Epidemiology},
  volume={32},
  pages={100319},
  year={2020},
  publisher={Elsevier}
}

@article{fuglstad2019constructing,
  title={Constructing priors that penalize the complexity of {Gaussian} random fields},
  author={Fuglstad, Geir-Arne and Simpson, Daniel and Lindgren, Finn and Rue, H{\aa}vard},
  journal={Journal of the American Statistical Association},
  volume={114},
  number={525},
  pages={445--452},
  year={2019},
  publisher={Taylor \& Francis}
}

@article{sorbye2014scaling,
  title={Scaling intrinsic Gaussian Markov random field priors in spatial modelling},
  author={S{\o}rbye, Sigrunn Holbek and Rue, H{\aa}vard},
  journal={Spatial Statistics},
  volume={8},
  pages={39--51},
  year={2014},
  publisher={Elsevier}
}

@article{besag1974,
	author = {Besag, Julian},
	year = {1974},
	title = {Spatial interaction and the statistical analysis of lattice systems},
	journal = {Journal of the Royal Statistical Society: Series B},
	volume = {36},
	number = {2},
	pages = {192--225}
}

@article{macnab2014identification,
  title={On identification in Bayesian disease mapping and ecological--spatial regression models},
  author={MacNab, Ying C},
  journal={Statistical Methods in Medical Research},
  volume={23},
  number={2},
  pages={134--155},
  year={2014},
  publisher={SAGE Publications Sage UK: London, England}
}

@article{bakka2018spatial,
  title={Spatial modeling with {R-INLA}: A review},
  author={Bakka, Haakon and Rue, H{\aa}vard and Fuglstad, Geir-Arne and Riebler, Andrea and Bolin, David and Illian, Janine and Krainski, Elias and Simpson, Daniel and Lindgren, Finn},
  journal={Wiley Interdisciplinary Reviews: Computational Statistics},
  volume={10},
  number={6},
  pages={e1443},
  year={2018},
  publisher={Wiley Online Library}
}

@article{gneiting2006geostatistical,
  title={Geostatistical space-time models, stationarity, separability, and full symmetry},
  author={Gneiting, Tilmann and Genton, Marc G and Guttorp, Peter},
  journal={Monographs on Statistics and Applied Probability},
  volume={107},
  pages={151},
  year={2006},
  publisher={Chapman \& Hall}
}

@article{lindgren2015bayesian,
  title={Bayesian spatial modelling with R-INLA},
  author={Lindgren, Finn and Rue, H{\aa}vard},
  journal={Journal of statistical software},
  volume={63},
  pages={1--25},
  year={2015}
}

@article{cayuela2020analisis,
  title={An{\'a}lisis espacial de la mortalidad por c{\'a}ncer de pulm{\'o}n en {Espa{\~n}a} con especial referencia a {Andaluc{\'\i}}a (2013-2017)},
  author={Cayuela, Lucia and Llanes, Adriano Asuero and Rodr{\'\i}guez-Dom{\'\i}nguez, Susana and Ch{\'a}vez-Conde, Manuel and Cayuela, Aurelio},
  journal={Open Respiratory Archives},
  volume={2},
  number={3},
  pages={107--112},
  year={2020},
  publisher={Elsevier}
}

@article{medin2025spatiotemporal,
  title={Spatiotemporal analysis of lung cancer mortality in {Spain}},
  author={Med{\'\i}n, Pablo and Leveau, Carlos M and Riancho, Javier and Santurt{\'u}n, Ana},
  journal={Cancer Epidemiology},
  volume={95},
  pages={102770},
  year={2025},
  publisher={Elsevier}
}

@article{jimenez2011mujeres,
  title={Mujeres, abandono de consumo de tabaco y barreras de g{\'e}nero},
  author={Jim{\'e}nez Rodrigo, Mar{\'\i}a Luisa},
  journal={Feminismo/s},
  volume={18},
  pages={39--65},
  year={2011},
  doi={http://dx.doi.org/10.14198/fem.2011.18.03},
  publisher={Universidad de Alicante. Centro de Estudios sobre la Mujer}
}

@article{bramajo2024lung,
  title={Lung cancer mortality trends among women across {Spain}: the role of birth cohorts in diverging regional patterns},
  author={Bramajo, Octavio},
  journal={Journal of Population Research},
  volume={41},
  number={1},
  pages={2},
  year={2024},
  publisher={Springer}
}

@article{clayton1987empirical,
  title={Empirical Bayes estimates of age-standardized relative risks for use in disease mapping},
  author={Clayton, David and Kaldor, John},
  journal={Biometrics},
  pages={671--681},
  year={1987},
  publisher={JSTOR}
}



%
\FloatBarrier
\label{lastmainpage}

\clearpage

\pagenumbering{arabic}
\setcounter{page}{1}

\renewcommand{\thefigure}{S\arabic{figure}}
\renewcommand{\theHfigure}{supp.\arabic{figure}}
\setcounter{figure}{0}

\renewcommand{\thetable}{S\arabic{table}}
\renewcommand{\theHtable}{supp.\arabic{table}}
\setcounter{table}{0}

\renewcommand{\theequation}{S\arabic{equation}}
\renewcommand{\theHequation}{supp.\arabic{equation}}
\setcounter{equation}{0}

\section*{Supplementary Material}

\setcounter{section}{0}
\setcounter{subsection}{0}
\setcounter{figure}{0}
\setcounter{table}{0}

\renewcommand{\thesection}{S\arabic{section}}
\renewcommand{\thesubsection}{S\arabic{section}.\arabic{subsection}}
\renewcommand{\thefigure}{S\arabic{figure}}
\renewcommand{\thetable}{S\arabic{table}}

This Supplementary Material provides additional results from the simulation study and the real-data analysis. Section~\ref{simulation_suppl} presents the estimated relative risk surfaces obtained with the SPDE models without the iid component, complementing the simulation results presented in the main paper.   SPDE models with the iid component are not including due to the large computing time to run the simulations.

Section~\ref{real_suppl} presents the estimated relative risks and posterior exceedance probabilities from the real data analysis for the 45--59 and 60--74 age groups. The corresponding results for individuals aged 75 years or older are presented in the main paper.

\section{Simulation study} \label{simulation_suppl}
\begin{figure}[H]
	\centering
	\includegraphics[width=\textwidth,
	height=0.55\textheight,
	keepaspectratio]{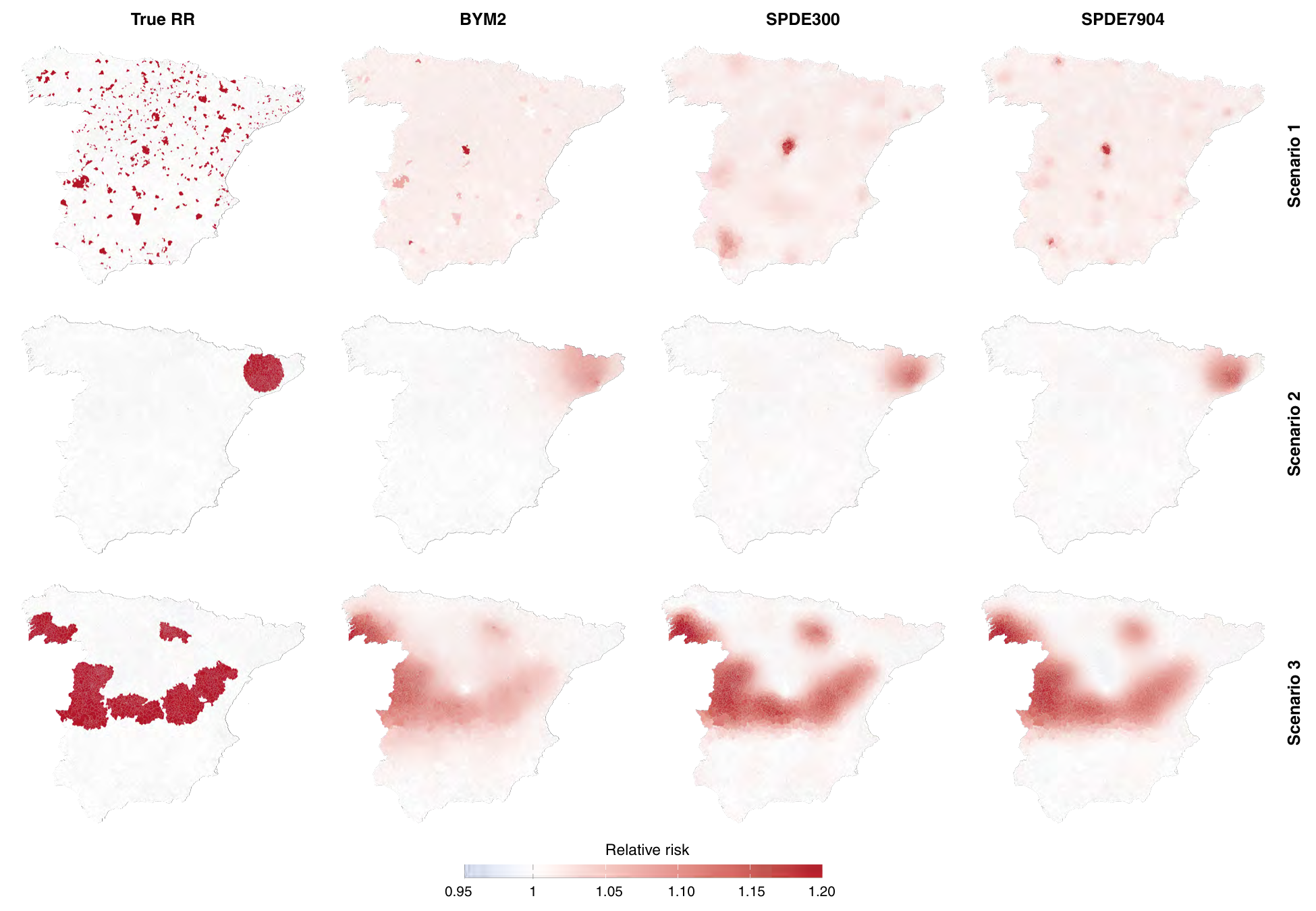}
	\caption{True and mean estimated relative risk surfaces for a true relative risk of 1.2 using the original expected counts. Rows correspond to the three simulation Scenarios. The first column shows the true relative risk, while the remaining columns show the mean municipal relative risk estimated across the 100 simulations.}
	\label{fig:simulation_rr_surfaces_noiid}
\end{figure}

Figure~\ref{fig:simulation_rr_surfaces_noiid} shows the true and the mean estimated relative risk surfaces obtained with the SPDE models without the iid component for a true relative risk of 1.2 using the original expected counts. These models were included in the simulation study to assess the effect of adding an area-specific unstructured random effect to the continuous spatial formulation. Overall, the spatial patterns are similar to those obtained with the iid formulations presented in the main paper. However, the numerical results showed that including the iid component generally improved the performance of SPDE7904, particularly in terms of specificity and false-positive detection. For this reason, the main paper focuses on the risk surfaces obtained with the iid formulations, while the corresponding results without the iid component are reported here.

Tables~\ref{tab:simulation_scenario1}, \ref{tab:simulation_scenario2} and \ref{tab:simulation_scenario3} report the complete simulation results for Scenarios~1--3 across all true relative risk and expected-count multiplying factors. The corresponding results for high-risk areas are presented in Tables~\ref{tab:simulation_risk_areas_s1},~\ref{tab:simulation_risk_areas_s2} and \ref{tab:simulation_risk_areas_s3}. These results complement the Tables presented in the main manuscript by showing how model performance changes as the amount of information increases.

Across the three Scenarios, increasing the expected counts generally reduces estimation errors, although the amount of improvement depends on the model and the spatial configuration of the high-risk areas. In Scenario~1, where high-risk municipalities are randomly scattered, improvements are relatively minor and less regular. For a true relative risk of 2 (Table~\ref{tab:simulation_scenario1}), for example, MARB decreases from 7.63\% to 4.84\% for BYM2 and from 10.70\% to 7.46\% for SPDE7904 as the expected-count multiplying factor increases from 1 to 10, while SPDE300 shows a less regular pattern and MARB remains around 10\% at the highest expected-count multiplying factor.

The effect of increasing expected counts is clearer in the clustered Scenarios~2 and~3. In Scenario~2, estimation errors generally decrease for all three models and sensitivity remains close to one over all configurations. For example, among high-risk municipalities with a true relative risk of 1.2 (Table~\ref{tab:simulation_risk_areas_s2}), MARB decreases from 9.84\% to 4.18\% for BYM2, from 9.27\% to 4.41\% for SPDE300 and from 8.02\% to 3.41\% for SPDE7904 as the expected counts increase from 1 to 10. Coverage within high-risk cluster also improves for three models.

Scenario~3 shows a similar improvement, although differences between SPDE300 and the other two models become more apparent with higher risk levels. For a true relative risk of 2 (Table~\ref{tab:simulation_scenario3}), RRMSE decrease substantially for BYM2 and SPDE7904. Within high-risk municipalities (Table~\ref{tab:simulation_risk_areas_s3}), SPDE7904 reduces MARB from 7.89\% to 4.73\% as the expected-count multiplying factor increases from 1 to 10. In contrast, SPDE300 shows an increase in MARB from 8.55\% to 10.04\% and a reduction in coverage from 0.788 to 0.574. This indicates that increasing the amount of information does not necessarily compensate for the loss of municipal-level spatial resolution associated with the aggregated SPDE model.

Overall, the high-risk municipality results show that global performance measures can hide poorer estimation in areas where the true risk is elevated. This is particularly evident in Scenario~1, where RRMSE and MARB remain substantially larger and coverage lower within high-risk municipalities than in the overall results. In the clustered scenarios, estimation within high-risk municipalities improves more clearly as the expected counts increase, with SPDE7904 generally benefiting most consistently from the additional information.
\begin{table}[!htbp]
	\centering
	\resizebox{\textwidth}{!}{
		\begin{talltblr}[
			caption = {Simulation results for Scenario 1, characterized by randomly scattered high-risk municipalities. RRMSE and MARB are expressed as percentages. Mean FP denotes the mean number of false-positive municipalities across simulations.},
			label = {tab:simulation_scenario1}
			]{
				colspec = {c c l r r r r r r},
				row{1} = {font=\bfseries},
				column{1-2} = {halign=c},
				column{3} = {halign=l},
				column{4-9} = {halign=r},
				rows = {font=\fontsize{7.4}{8.6}\selectfont},
				rowsep = 1.5pt
			}
			
			\toprule
			RR & Exp. counts & Model & RRMSE (\%) & MARB (\%) &
			95\% Coverage & Sensitivity & Specificity & Mean FP \\
			\midrule
			
			1.2 & $\times1$ & BYM2
			& 2.55 & 2.25 & 0.937 & 0.041 & 0.994 & 42.0 \\
			1.2 & $\times1$ & SPDE300
			& 3.37 & 2.51 & 0.921 & 0.169 & 0.854 & 1077.0 \\
			1.2 & $\times1$ & SPDE7904
			& 3.13 & 2.32 & 0.935 & 0.099 & 0.946 & 401.1 \\
			
			1.2 & $\times2$ & BYM2
			& 2.56 & 2.24 & 0.937 & 0.076 & 0.989 & 78.6 \\
			1.2 & $\times2$ & SPDE300
			& 3.29 & 2.46 & 0.914 & 0.225 & 0.810 & 1408.87 \\
			1.2 & $\times2$ & SPDE7904
			& 3.13 & 2.29 & 0.934 & 0.138 & 0.935 & 480.61 \\
			
			1.2 & $\times5$ & BYM2
			& 2.57 & 2.14 & 0.938 & 0.159 & 0.978 & 162.8 \\
			1.2 & $\times5$ & SPDE300
			& 3.23 & 2.39 & 0.895 & 0.309 & 0.749 & 1861.05 \\
			1.2 & $\times5$ & SPDE7904
			& 3.18 & 2.21 & 0.933 & 0.226 & 0.912 & 645.61 \\
			
			1.2 & $\times10$ & BYM2
			& 2.60 & 2.05 & 0.938 & 0.260 & 0.963 & 275.3 \\
			1.2 & $\times10$ & SPDE300
			& 3.18 & 2.39 & 0.869 & 0.378 & 0.702 & 2204.86 \\
			1.2 & $\times10$ & SPDE7904
			& 3.25 & 2.16 & 0.933 & 0.324 & 0.889 & 820.84 \\
			
			\midrule
			
			1.5 & $\times1$ & BYM2
			& 5.81 & 4.72 & 0.938 & 0.174 & 0.977 & 172.7 \\
			1.5 & $\times1$ & SPDE300
			& 7.53 & 5.75 & 0.885 & 0.342 & 0.723 & 2053.0 \\
			1.5 & $\times1$ & SPDE7904
			& 7.58 & 5.51 & 0.935 & 0.260 & 0.901 & 731.3 \\
			
			1.5 & $\times2$ & BYM2
			& 5.86 & 4.48 & 0.939 & 0.283 & 0.962 & 277.9 \\
			1.5 & $\times2$ & SPDE300
			& 7.33 & 5.66 & 0.858 & 0.403 & 0.679 & 2372.4 \\
			1.5 & $\times2$ & SPDE7904
			& 7.81 & 5.43 & 0.935 & 0.363 & 0.877 & 908.46 \\
			
			1.5 & $\times5$ & BYM2
			& 6.12 & 4.24 & 0.940 & 0.485 & 0.929 & 529.2 \\
			1.5 & $\times5$ & SPDE300
			& 7.12 & 5.66 & 0.812 & 0.492 & 0.626 & 2768.0 \\
			1.5 & $\times5$ & SPDE7904
			& 8.17 & 5.14 & 0.936 & 0.550 & 0.837 & 1204.07 \\
			
			1.5 & $\times10$ & BYM2
			& 6.09 & 3.80 & 0.941 & 0.655 & 0.899 & 746.6 \\
			1.5 & $\times10$ & SPDE300
			& 6.96 & 5.73 & 0.769 & 0.549 & 0.592 & 3020.36 \\
			1.5 & $\times10$ & SPDE7904
			& 8.15 & 4.66 & 0.937 & 0.699 & 0.809 & 1410.06 \\
			
			\midrule
			
			2.0 & $\times1$ & BYM2
			& 10.70 & 7.63 & 0.942 & 0.400 & 0.948 & 383.7 \\
			2.0 & $\times1$ & SPDE300
			& 13.4 & 10.70 & 0.820 & 0.481 & 0.626 & 2765.2 \\
			2.0 & $\times1$ & SPDE7904
			& 16.01 & 11.40 & 0.941 & 0.505 & 0.836 & 1217.6 \\
			
			2.0 & $\times2$ & BYM2
			& 10.80 & 6.95 & 0.944 & 0.560 & 0.924 & 564.6 \\
			2.0 & $\times2$ & SPDE300
			& 6.97 & 5.73 & 0.769 & 0.550 & 0.592 & 3021.75 \\
			2.0 & $\times2$ & SPDE7904
			& 16.22 & 10.82 & 0.941 & 0.652 & 0.803 & 1456.44 \\
			
			2.0 & $\times5$ & BYM2
			& 10.53 & 5.83 & 0.946 & 0.767 & 0.885 & 850.0 \\
			2.0 & $\times5$ & SPDE300
			& 12.59 & 10.83 & 0.711 & 0.651 & 0.543 & 3380.51 \\
			2.0 & $\times5$ & SPDE7904
			& 15.04 & 9.14 & 0.936 & 0.818 & 0.768 & 1714.54 \\
			
			2.0 & $\times10$ & BYM2
			& 9.99 & 4.84 & 0.948 & 0.885 & 0.856 & 1066.3 \\
			2.0 & $\times10$ & SPDE300
			& 12.29 & 10.95 & 0.657 & 0.656 & 0.518 & 3565.56 \\
			2.0 & $\times10$ & SPDE7904
			& 13.44 & 7.46 & 0.930 & 0.904 & 0.751 & 1846.46 \\
			
			\bottomrule
			
		\end{talltblr}
	}
\end{table}

\begin{table}[!htbp]
	\centering
	\resizebox{\textwidth}{!}{
		\begin{talltblr}[
			caption = {Simulation results for Scenario 2, characterized by a single contiguous high-risk cluster. RRMSE and MARB are expressed as percentages. Mean FP denotes the mean number of false-positive municipalities across simulations.},
			label = {tab:simulation_scenario2}
			]{
				colspec = {c c l r r r r r r},
				row{1} = {font=\bfseries},
				column{1-2} = {halign=c},
				column{3} = {halign=l},
				column{4-9} = {halign=r},
				rows = {font=\fontsize{7.4}{8.6}\selectfont},
				rowsep = 1.5pt
			}
			
			\toprule
			RR & Exp. counts & Model & RRMSE (\%) & MARB (\%) &
			95\% Coverage & Sensitivity & Specificity & Mean FP \\
			\midrule
			
			1.2 & $\times1$ & BYM2
			& 1.93 & 1.02 & 0.955 & 0.993 & 0.945 & 408.2 \\
			1.2 & $\times1$ & SPDE300
			& 2.14 & 0.90 & 0.949 & 0.969 & 0.916 & 626.03 \\
			1.2 & $\times1$ & SPDE7904
			& 2.21 & 0.81 & 0.954 & 0.979 & 0.918 & 610.7 \\
			
			1.2 & $\times2$ & BYM2
			& 1.85 & 0.83 & 0.964 & 0.999 & 0.946 & 401.8 \\
			1.2 & $\times2$ & SPDE300
			& 2.03 & 0.72 & 0.952 & 0.989 & 0.903 & 723.18 \\
			1.2 & $\times2$ & SPDE7904
			& 2.06 & 0.67 & 0.955 & 0.997 & 0.905 & 707.87 \\
			
			1.2 & $\times5$ & BYM2
			& 1.64 & 0.65 & 0.969 & 1.000 & 0.947 & 398.8 \\
			1.2 & $\times5$ & SPDE300
			& 1.80 & 0.57 & 0.952 & 0.998 & 0.887 & 832.33 \\
			1.2 & $\times5$ & SPDE7904
			& 1.79 & 0.52 & 0.956 & 0.999 & 0.894 & 791.45 \\
			
			1.2 & $\times10$ & BYM2
			& 1.52 & 0.55 & 0.972 & 1.000 & 0.944 & 416.0 \\
			1.2 & $\times10$ & SPDE300
			& 1.62 & 0.48 & 0.952 & 0.998 & 0.998 & 902.99 \\
			1.2 & $\times10$ & SPDE7904
			& 1.62 & 0.45 & 0.958 & 1.000 & 0.884 & 866.81 \\
			
			\midrule
			
			1.5 & $\times1$ & BYM2
			& 3.73 & 1.41 & 0.975 & 1.000 & 0.947 & 395.5 \\
			1.5 & $\times1$ & SPDE300
			& 4.08 & 1.31 & 0.955 & 0.998 & 0.882 & 879.8 \\
			1.5 & $\times1$ & SPDE7904
			& 4.10 & 1.18 & 0.964 & 0.960 & 0.891 & 813.4 \\
			
			1.5 & $\times2$ & BYM2
			& 3.36 & 1.19 & 0.974 & 1.000 & 0.947 & 394.5 \\
			1.5 & $\times2$ & SPDE300
			& 3.60 & 1.13 & 0.953 & 0.999 & 0.877 & 918.53 \\
			1.5 & $\times2$ & SPDE7904
			& 3.62 & 1.03 & 0.959 & 1.000 & 0.885 & 859.37 \\
			
			1.5 & $\times5$ & BYM2
			& 2.96 & 0.93 & 0.975 & 1.000 & 0.944 & 420.5 \\
			1.5 & $\times5$ & SPDE300
			& 3.10 & 0.945 & 0.950 & 0.999 & 0.873 & 946.86 \\
			1.5 & $\times5$ & SPDE7904
			& 3.13 & 0.82 & 0.962 & 1.000 & 0.880 & 895.96 \\
			
			1.5 & $\times10$ & BYM2
			& 2.74 & 0.78 & 0.975 & 1.000 & 0.937 & 471.8 \\
			1.5 & $\times10$ & SPDE300
			& 2.76 & 0.89 & 0.943 & 0.998 & 0.866 & 1000.22 \\
			1.5 & $\times10$ & SPDE7904
			& 2.85 & 0.72 & 0.964 & 1.000 & 0.873 & 951.8 \\
			
			\midrule
			
			2.0 & $\times1$ & BYM2
			& 5.54 & 1.86 & 0.977 & 1.000 & 0.947 & 399.7 \\
			2.0 & $\times1$ & SPDE300
			& 5.98 & 2.02 & 0.953 & 0.999 & 0.869 & 979.3 \\
			2.0 & $\times1$ & SPDE7904
			& 5.99 & 1.70 & 0.963 & 1.000 & 0.883 & 874.6 \\
			
			2.0 & $\times2$ & BYM2
			& 5.10 & 1.56 & 0.977 & 1.000 & 0.943 & 428.1 \\
			2.0 & $\times2$ & SPDE300
			& 5.30 & 1.56 & 0.948 & 0.995 & 0.882 & 879.2 \\
			2.0 & $\times2$ & SPDE7904
			& 5.44 & 1.46 & 0.965 & 1.000 & 0.878 & 909.51 \\
			
			2.0 & $\times5$ & BYM2
			& 4.52 & 1.21 & 0.977 & 1.000 & 0.935 & 486.7 \\
			2.0 & $\times5$ & SPDE300
			& 4.45 & 1.48 & 0.937 & 0.993 & 0.880 & 899.9 \\
			2.0 & $\times5$ & SPDE7904
			& 4.69 & 1.20 & 0.967 & 1.000 & 0.874 & 942.56 \\
			
			2.0 & $\times10$ & BYM2
			& 4.15 & 1.01 & 0.977 & 1.000 & 0.925 & 557.1 \\
			2.0 & $\times10$ & SPDE300
			& 3.94 & 1.43 & 0.924 & 0.991 & 0.875 & 937.1 \\
			2.0 & $\times10$ & SPDE7904
			& 4.29 & 1.00 & 0.969 & 1.000 & 0.869 & 981.89 \\
			
			\bottomrule
			
		\end{talltblr}
	}
\end{table}

\begin{table}[!htbp]
	\centering
	\resizebox{\textwidth}{!}{
		\begin{talltblr}[
			caption = {Simulation results for Scenario 3, characterized by five non-overlapping high-risk clusters. RRMSE and MARB are expressed as percentages. Mean FP denotes the mean number of false-positive municipalities across simulations.},
			label = {tab:simulation_scenario3}
			]{
				colspec = {c c l r r r r r r},
				row{1} = {font=\bfseries},
				column{1-2} = {halign=c},
				column{3} = {halign=l},
				column{4-9} = {halign=r},
				rows = {font=\fontsize{7.4}{8.6}\selectfont},
				rowsep = 1.5pt
			}
			
			\toprule
			RR & Exp. counts & Model & RRMSE (\%) & MARB (\%) &
			95\% Coverage & Sensitivity & Specificity & Mean FP \\
			\midrule
			
			1.2 & $\times1$ & BYM2
			& 4.10 & 3.22 & 0.910 & 0.863 & 0.873 & 805.7 \\
			1.2 & $\times1$ & SPDE300
			& 4.06 & 2.37 & 0.912 & 0.974 & 0.770 & 1455.7 \\
			1.2 & $\times1$ & SPDE7904
			& 3.94 & 2.55 & 0.865 & 0.981 & 0.735 & 1676.0 \\
			
			1.2 & $\times2$ & BYM2
			& 3.72 & 2.76 & 0.923 & 0.959 & 0.863 & 864.9 \\
			1.2 & $\times2$ & SPDE300
			& 3.62 & 1.89 & 0.923 & 0.990 & 0.788 & 1343.13 \\
			1.2 & $\times2$ & SPDE7904
			& 3.50 & 2.13 & 0.872 & 0.994 & 0.750 & 1584.66 \\
			
			1.2 & $\times5$ & BYM2
			& 3.25 & 2.22 & 0.936 & 0.994 & 0.854 & 927.8 \\
			1.2 & $\times5$ & SPDE300
			& 3.09 & 1.49 & 0.921 & 0.997 & 0.797 & 1283.3 \\
			1.2 & $\times5$ & SPDE7904
			& 2.96 & 1.72 & 0.874 & 0.999 & 0.759 & 1528.87 \\
			
			1.2 & $\times10$ & BYM2
			& 2.91 & 1.87 & 0.942 & 0.998 & 0.853 & 933.9 \\
			1.2 & $\times10$ & SPDE300
			& 2.75 & 1.35 & 0.915 & 0.998 & 0.804 & 1213.5 \\
			1.2 & $\times10$ & SPDE7904
			& 2.59 & 1.45 & 0.881 & 1.000 & 0.773 & 1440.04 \\
			
			\midrule
			
			1.5 & $\times1$ & BYM2
			& 7.14 & 4.73 & 0.948 & 0.994 & 0.860 & 888.7 \\
			1.5 & $\times1$ & SPDE300
			& 6.94 & 3.44 & 0.922 & 0.998 & 0.794 & 1305.3 \\
			1.5 & $\times1$ & SPDE7904
			& 6.57 & 3.81 & 0.880 & 0.999 & 0.759 & 1524.0 \\
			
			1.5 & $\times2$ & BYM2
			& 6.42 & 3.95 & 0.949 & 0.998 & 0.861 & 882.0 \\
			1.5 & $\times2$ & SPDE300
			& 6.14 & 3.09 & 0.913 & 0.999 & 0.810 & 1205.19 \\
			1.5 & $\times2$ & SPDE7904
			& 5.77 & 3.22 & 0.884 & 1.000 & 0.775 & 1426.96 \\
			
			1.5 & $\times5$ & BYM2
			& 5.55 & 3.02 & 0.948 & 1.000 & 0.863 & 866.0 \\
			1.5 & $\times5$ & SPDE300
			& 5.35 & 2.91 & 0.892 & 0.998 & 0.820 & 1143.59 \\
			1.5 & $\times5$ & SPDE7904
			& 4.91 & 2.59 & 0.893 & 1.000 & 0.800 & 1295.32 \\
			
			1.5 & $\times10$ & BYM2
			& 4.98 & 2.43 & 0.948 & 1.000 & 0.862 & 875.3 \\
			1.5 & $\times10$ & SPDE300
			& 4.84 & 2.96 & 0.868 & 0.997 & 0.827 & 1096.19 \\
			1.5 & $\times10$ & SPDE7904
			& 4.34 & 2.16 & 0.903 & 1.000 & 0.808 & 1215.85 \\
			
			\midrule
			
			2.0 & $\times1$ & BYM2
			& 10.40 & 5.97 & 0.954 & 0.999 & 0.865 & 854.0 \\
			2.0 & $\times1$ & SPDE300
			& 10.01 & 5.46 & 0.901 & 0.999 & 0.809 & 1213.4 \\
			2.0 & $\times1$ & SPDE7904
			& 9.36 & 5.05 & 0.891 & 1.000 & 0.784 & 1372.3 \\
			
			2.0 & $\times2$ & BYM2
			& 9.28 & 4.82 & 0.951 & 1.000 & 0.865 & 853.5 \\
			2.0 & $\times2$ & SPDE300
			& 9.08 & 5.30 & 0.883 & 0.998 & 0.820 & 1138.17 \\
			2.0 & $\times2$ & SPDE7904
			& 8.25 & 4.25 & 0.898 & 1.000 & 0.800 & 1267.66 \\
			
			2.0 & $\times5$ & BYM2
			& 8.05 & 3.52 & 0.951 & 1.000 & 0.865 & 858.1 \\
			2.0 & $\times5$ & SPDE300
			& 8.21 & 5.61 & 0.843 & 0.996 & 0.825 & 1106.8 \\
			2.0 & $\times5$ & SPDE7904
			& 7.06 & 3.33 & 0.912 & 1.000 & 0.813 & 1182.89 \\
			
			2.0 & $\times10$ & BYM2
			& 7.23 & 2.75 & 0.951 & 1.000 & 0.859 & 895.3 \\
			2.0 & $\times10$ & SPDE300
			& 7.72 & 5.86 & 0.807 & 0.994 & 0.824 & 1115.92 \\
			2.0 & $\times10$ & SPDE7904
			& 6.32 & 2.75 & 0.919 & 1.000 & 0.820 & 1142.07 \\
			
			\bottomrule
			
		\end{talltblr}
	}
\end{table}


\begin{table}[!htbp]
	\centering
	\resizebox{\width}{3.85in}{
		\begin{talltblr}[
			caption = {Simulation results for high-risk municipalities in Scenario~1. RRMSE and MARB are expressed as percentages.},
			label = {tab:simulation_risk_areas_s1}
			]{
				colspec = {c c l r r r},
				row{1} = {font=\bfseries},
				column{1-2} = {halign=c},
				column{3} = {halign=l},
				column{4-6} = {halign=r},
				rows = {font=\fontsize{7.4}{8.6}\selectfont},
				rowsep = 1.5pt
			}
			
			\toprule
			RR & Expected counts & Model & RRMSE (\%) & MARB (\%) & 95\% Coverage \\
			\midrule
			
			1.2 & $\times 1$  & BYM2     & 15.10 & 15.10 & 0.012\\
			1.2 & $\times 1$  & SPDE300  & 15.30 & 15.20 & 0.005\\
			1.2 & $\times 1$  & SPDE7904 & 15.20 & 15.10 & 0.016\\
			
			1.2 & $\times 2$  & BYM2     & 14.87 & 14.83 & 0.014\\
			1.2 & $\times 2$  & SPDE300  & 15.27 & 15.20 & 0.004\\
			1.2 & $\times 2$  & SPDE7904 & 14.95 & 14.86 & 0.017\\
			
			1.2 & $\times 5$  & BYM2     & 14.40 & 14.34 & 0.022\\
			1.2 & $\times 5$  & SPDE300  & 15.24 & 15.17 & 0.002\\
			1.2 & $\times 5$  & SPDE7904 & 14.52 & 14.41 & 0.026\\
			
			1.2 & $\times 10$ & BYM2     & 13.83 & 13.75 & 0.034\\
			1.2 & $\times 10$ & SPDE300  & 15.16 & 15.10 & 0.002\\
			1.2 & $\times 10$ & SPDE7904 & 13.93 & 13.79 & 0.043\\
			
			\addlinespace[0.25em]
			\midrule
			
			1.5 & $\times 1$  & BYM2     & 28.70 & 28.50 & 0.036\\
			1.5 & $\times 1$  & SPDE300  & 30.30 & 30.10 & 0.003\\
			1.5 & $\times 1$  & SPDE7904 & 28.50 & 28.20 & 0.05\\
			
			1.5 & $\times 2$  & BYM2     & 27.45 & 27.25 & 0.054\\
			1.5 & $\times 2$  & SPDE300  & 30.21 & 30.10 & 0.002\\
			1.5 & $\times 2$  & SPDE7904 & 27.10 & 26.76 & 0.095\\
			
			1.5 & $\times 5$  & BYM2     & 24.72 & 24.39 & 0.107\\
			1.5 & $\times 5$  & SPDE300  & 30.00 & 29.91 & 0.001\\
			1.5 & $\times 5$  & SPDE7904 & 24.10 & 23.50 & 0.214\\
			
			1.5 & $\times 10$ & BYM2     & 22.01 & 21.55 & 0.173\\
			1.5 & $\times 10$ & SPDE300  & 29.84 & 29.78 & 0.0003\\
			1.5 & $\times 10$ & SPDE7904 & 20.94 & 20.09 & 0.339\\
			
			\addlinespace[0.25em]
			\midrule
			
			2.0 & $\times 1$  & BYM2     & 38.90 & 38.30 & 0.121\\
			2.0 & $\times 1$  & SPDE300  & 44.90 & 44.80 & 0.002\\
			2.0 & $\times 1$  & SPDE7904 & 36.80 & 35.70 & 0.291\\
			
			2.0 & $\times 2$  & BYM2     & 35.25 & 34.49 & 0.188\\
			2.0 & $\times 2$  & SPDE300  & 29.85 & 29.78 & 0.0003\\
			2.0 & $\times 2$  & SPDE7904 & 32.40 & 30.85 & 0.426\\
			
			2.0 & $\times 5$  & BYM2     & 29.31 & 28.22 & 0.301\\
			2.0 & $\times 5$  & SPDE300  & 44.50 & 44.42 & 0.0006\\
			2.0 & $\times 5$  & SPDE7904 & 26.30 & 24.01 & 0.550\\
			
			2.0 & $\times 10$ & BYM2     & 24.02 & 22.69 & 0.413\\
			2.0 & $\times 10$ & SPDE300  & 44.40 & 44.30 & 0.0005\\
			2.0 & $\times 10$ & SPDE7904 & 21.40 & 18.87 & 0.620\\
			
			\bottomrule
			
		\end{talltblr}
	}
\end{table}


\begin{table}[!htbp]
	\centering
	\resizebox{\width}{3.85in}{
		\begin{talltblr}[
			caption = {Simulation results for high-risk municipalities in Scenario~2. RRMSE and MARB are expressed as percentages.},
			label = {tab:simulation_risk_areas_s2}
			]{
				colspec = {c c l r r r},
				row{1} = {font=\bfseries},
				column{1-2} = {halign=c},
				column{3} = {halign=l},
				column{4-6} = {halign=r},
				rows = {font=\fontsize{7.4}{8.6}\selectfont},
				rowsep = 1.5pt
			}
			
			\toprule
			RR & Expected counts & Model & RRMSE (\%) & MARB (\%) & 95\% Coverage \\
			\midrule
			
			1.2 & $\times 1$  & BYM2     & 10.00 & 9.84 & 0.216\\
			1.2 & $\times 1$  & SPDE300  & 9.59  & 9.27 & 0.230\\
			1.2 & $\times 1$  & SPDE7904 & 8.43  & 8.05 & 0.315\\
			
			1.2 & $\times 2$  & BYM2     & 7.73 & 7.56 & 0.433\\
			1.2 & $\times 2$  & SPDE300  & 7.66 & 7.29 & 0.342\\
			1.2 & $\times 2$  & SPDE7904 & 6.62 & 6.21 & 0.424\\
			
			1.2 & $\times 5$  & BYM2     & 5.70 & 5.55 & 0.618\\
			1.2 & $\times 5$  & SPDE300  & 5.82 & 5.42 & 0.445\\
			1.2 & $\times 5$  & SPDE7904 & 4.86 & 4.49 & 0.527\\
			
			1.2 & $\times 10$ & BYM2     & 4.41 & 4.18 & 0.719\\
			1.2 & $\times 10$ & SPDE300  & 4.86 & 4.41 & 0.513\\
			1.2 & $\times 10$ & SPDE7904 & 3.84 & 3.41 & 0.605\\
			
			\addlinespace[0.25em]
			\midrule
			
			1.5 & $\times 1$  & BYM2     & 10.60 & 10.10 & 0.724\\
			1.5 & $\times 1$  & SPDE300  & 11.30 & 10.20 & 0.509\\
			1.5 & $\times 1$  & SPDE7904 & 9.47  & 8.48  & 0.591\\
			
			1.5 & $\times 2$  & BYM2     & 8.62 & 8.02 & 0.757\\
			1.5 & $\times 2$  & SPDE300  & 9.59 & 8.57 & 0.547\\
			1.5 & $\times 2$  & SPDE7904 & 7.58 & 6.52 & 0.640\\
			
			1.5 & $\times 5$  & BYM2     & 6.43 & 5.64 & 0.806\\
			1.5 & $\times 5$  & SPDE300  & 8.09 & 7.16 & 0.574\\
			1.5 & $\times 5$  & SPDE7904 & 5.71 & 4.64 & 0.713\\
			
			1.5 & $\times 10$ & BYM2     & 5.25 & 4.30 & 0.826\\
			1.5 & $\times 10$ & SPDE300  & 7.29 & 6.43 & 0.570\\
			1.5 & $\times 10$ & SPDE7904 & 4.78 & 3.75 & 0.756\\
			
			\addlinespace[0.25em]
			\midrule
			
			2.0 & $\times 1$  & BYM2     & 10.80 & 9.27  & 0.837\\
			2.0 & $\times 1$  & SPDE300  & 13.20 & 11.40 & 0.599\\
			2.0 & $\times 1$  & SPDE7904 & 9.83  & 7.67  & 0.729\\
			
			2.0 & $\times 2$  & BYM2     & 8.95  & 7.19  & 0.847\\
			2.0 & $\times 2$  & SPDE300  & 12.93 & 11.31 & 0.564\\
			2.0 & $\times 2$  & SPDE7904 & 8.19  & 6.32  & 0.764\\
			
			2.0 & $\times 5$  & BYM2     & 7.01  & 4.96  & 0.864\\
			2.0 & $\times 5$  & SPDE300  & 11.77 & 10.62 & 0.526\\
			2.0 & $\times 5$  & SPDE7904 & 6.62  & 4.92  & 0.799\\
			
			2.0 & $\times 10$ & BYM2     & 6.03  & 3.90  & 0.867\\
			2.0 & $\times 10$ & SPDE300  & 11.51 & 10.71 & 0.449\\
			2.0 & $\times 10$ & SPDE7904 & 5.77  & 3.99  & 0.818\\
			
			\bottomrule
			
		\end{talltblr}
	}
\end{table}


\begin{table}[!htbp]
	\centering
	\resizebox{\width}{3.85in}{
		\begin{talltblr}[
			caption = {Simulation results for high-risk municipalities in Scenario~3. RRMSE and MARB are expressed as percentages.},
			label = {tab:simulation_risk_areas_s3}
			]{
				colspec = {c c l r r r},
				row{1} = {font=\bfseries},
				column{1-2} = {halign=c},
				column{3} = {halign=l},
				column{4-6} = {halign=r},
				rows = {font=\fontsize{7.4}{8.6}\selectfont},
				rowsep = 1.5pt
			}
			
			\toprule
			RR & Expected counts & Model & RRMSE (\%) & MARB (\%) & 95\% Coverage \\
			\midrule
			
			1.2 & $\times 1$  & BYM2     & 9.01 & 8.74 & 0.576\\
			1.2 & $\times 1$  & SPDE300  & 6.41 & 5.51 & 0.751\\
			1.2 & $\times 1$  & SPDE7904 & 6.86 & 6.24 & 0.587\\
			
			1.2 & $\times 2$  & BYM2     & 7.61 & 7.26 & 0.664\\
			1.2 & $\times 2$  & SPDE300  & 5.21 & 4.04 & 0.812\\
			1.2 & $\times 2$  & SPDE7904 & 5.67 & 4.95 & 0.648\\
			
			1.2 & $\times 5$  & BYM2     & 5.97 & 5.54 & 0.766\\
			1.2 & $\times 5$  & SPDE300  & 4.12 & 2.92 & 0.834\\
			1.2 & $\times 5$  & SPDE7904 & 4.46 & 3.72 & 0.694\\
			
			1.2 & $\times 10$ & BYM2     & 4.86 & 4.32 & 0.821\\
			1.2 & $\times 10$ & SPDE300  & 3.70 & 2.61 & 0.816\\
			1.2 & $\times 10$ & SPDE7904 & 3.78 & 3.03 & 0.723\\
			
			\addlinespace[0.25em]
			\midrule
			
			1.5 & $\times 1$  & BYM2     & 11.40 & 10.30 & 0.826\\
			1.5 & $\times 1$  & SPDE300  & 8.38  & 5.75  & 0.845\\
			1.5 & $\times 1$  & SPDE7904 & 8.73  & 7.02  & 0.732\\
			
			1.5 & $\times 2$  & BYM2     & 9.70 & 8.37 & 0.852\\
			1.5 & $\times 2$  & SPDE300  & 7.74 & 5.43 & 0.815\\
			1.5 & $\times 2$  & SPDE7904 & 7.65 & 5.93 & 0.742\\
			
			1.5 & $\times 5$  & BYM2     & 7.81 & 6.21 & 0.873\\
			1.5 & $\times 5$  & SPDE300  & 7.20 & 5.47 & 0.750\\
			1.5 & $\times 5$  & SPDE7904 & 6.53 & 4.89 & 0.764\\
			
			1.5 & $\times 10$ & BYM2     & 6.62 & 4.83 & 0.879\\
			1.5 & $\times 10$ & SPDE300  & 6.98 & 5.76 & 0.683\\
			1.5 & $\times 10$ & SPDE7904 & 5.78 & 4.22 & 0.782\\
			
			\addlinespace[0.25em]
			\midrule
			
			2.0 & $\times 1$  & BYM2     & 13.10 & 10.20 & 0.891\\
			2.0 & $\times 1$  & SPDE300  & 11.80 & 8.55  & 0.788\\
			2.0 & $\times 1$  & SPDE7904 & 10.80 & 7.99  & 0.777\\
			
			2.0 & $\times 2$  & BYM2     & 11.38 & 8.19 & 0.892\\
			2.0 & $\times 2$  & SPDE300  & 11.38 & 8.84 & 0.725\\
			2.0 & $\times 2$  & SPDE7904 & 9.73  & 6.94 & 0.788\\
			
			2.0 & $\times 5$  & BYM2     & 9.51  & 5.88 & 0.898\\
			2.0 & $\times 5$  & SPDE300  & 11.10 & 9.54 & 0.634\\
			2.0 & $\times 5$  & SPDE7904 & 8.35  & 5.64 & 0.812\\
			
			2.0 & $\times 10$ & BYM2     & 8.30  & 4.54  & 0.902\\
			2.0 & $\times 10$ & SPDE300  & 10.98 & 10.04 & 0.574\\
			2.0 & $\times 10$ & SPDE7904 & 7.41  & 4.73  & 0.826\\
			
			\bottomrule
			
		\end{talltblr}
	}
\end{table}

\section{Real-data analysis} \label{real_suppl}

\subsection{Analysis of younger age groups}

Tables~\ref{tab:model_comparison_45_59} and~\ref{tab:model_comparison_60_74} show the model comparison for the 45--59 and 60--74 age groups, respectively. The results are consistent with those obtained for the 75+ age group presented in the main paper. For SPDE7904, including the iid component reduces WAIC in both sexes and age groups, bringing the values closer to those obtained with BYM2, while mean log-scores remain very similar. The iid formulation also reduces the percentage of municipalities classified as high risk compared with SPDE7904 without the iid component. For SPDE300, the differences between the two formulations are smaller, although the iid model generally yields lower WAIC values.

\clearpage{}

\begin{table}[!htpb]
	\centering
	\begin{talltblr}[
		caption = {Model comparison for the 45-59 age group by sex.},
		label = {tab:model_comparison_45_59}
		]{
			colspec = {l l r r r},
			row{1} = {font=\bfseries},
			column{1} = {halign=l},
			column{2} = {halign=l},
			column{3-5} = {halign=r}
		}
		\toprule
		Sex & Model & WAIC & Mean Log-score & \% of high-risk areas \\
		\midrule
		\SetCell[r=3]{m} Males
		& BYM2     & 18122.05 & 1.15 & 7.87  \\
		\cline[dashed]{2-5}
		& SPDE7904 & 18289.35 & 1.16 & 12.53 \\
		& SPDE7904 (iid) & 18137.87 & 1.15 & 9.83 \\
		\cline[dashed]{2-5}
		& SPDE300  & 2022.69  & 3.48 & 21.07 \\
		& SPDE300 (iid) & 1993.57  & 3.44 & 21.01 \\
		\midrule
		\SetCell[r=3]{m} Females
		& BYM2     & 10658.86 & 0.68 & 5.43 \\
		\cline[dashed]{2-5}
		& SPDE7904 & 10746.07 & 0.68 & 9.92 \\
		& SPDE7904 (iid) & 10657.28 & 0.68 & 6.85 \\
		\cline[dashed]{2-5}
		& SPDE300  & 1675.64  & 2.86 & 13.82 \\
		& SPDE300 (iid) & 1655.31  & 2.84 & 12.44 \\
		
		\bottomrule
	\end{talltblr}
\end{table}

\begin{table}[!htpb]
	\centering
	\begin{talltblr}[
		caption = {Model comparison for the 60-74 age group by sex.},
		label = {tab:model_comparison_60_74}
		]{
			colspec = {l l r r r},
			row{1} = {font=\bfseries},
			column{1} = {halign=l},
			column{2} = {halign=l},
			column{3-5} = {halign=r}
		}
		\toprule
		Sex & Model & WAIC & Mean Log-score & \% of high-risk areas \\
		\midrule
		\SetCell[r=3]{m} Males
		& BYM2     & 26570.87 & 1.70 & 10.77  \\
		\cline[dashed]{2-5}
		& SPDE7904 & 26770.27 & 1.70 & 16.75 \\
		& SPDE7904 (iid) & 26561.44 & 1.69 & 13.61 \\
		\cline[dashed]{2-5}
		& SPDE300  & 2442.80  & 4.32 & 22.86 \\
		& SPDE300 (iid) & 2407.35  & 4.27 & 22.63 \\
		\midrule
		\SetCell[r=3]{m} Females
		& BYM2     & 13280.86 & 0.86 & 2.44 \\
		\cline[dashed]{2-5}
		& SPDE7904 & 13361.35 & 0.85 & 4.89 \\
		& SPDE7904 (iid) & 13280.03 & 0.85 & 3.43 \\
		\cline[dashed]{2-5}
		& SPDE300  & 1822.17  & 3.21 & 11.88 \\
		& SPDE300 (iid) & 1815.98  & 3.23 & 10.82 \\
		
		\bottomrule
	\end{talltblr}
\end{table}

\subsection{Additional maps for the younger age groups}

\begin{figure}[H]
	\centering
	\includegraphics[
	width=\textwidth,
	height=0.80\textheight,
	keepaspectratio
	]{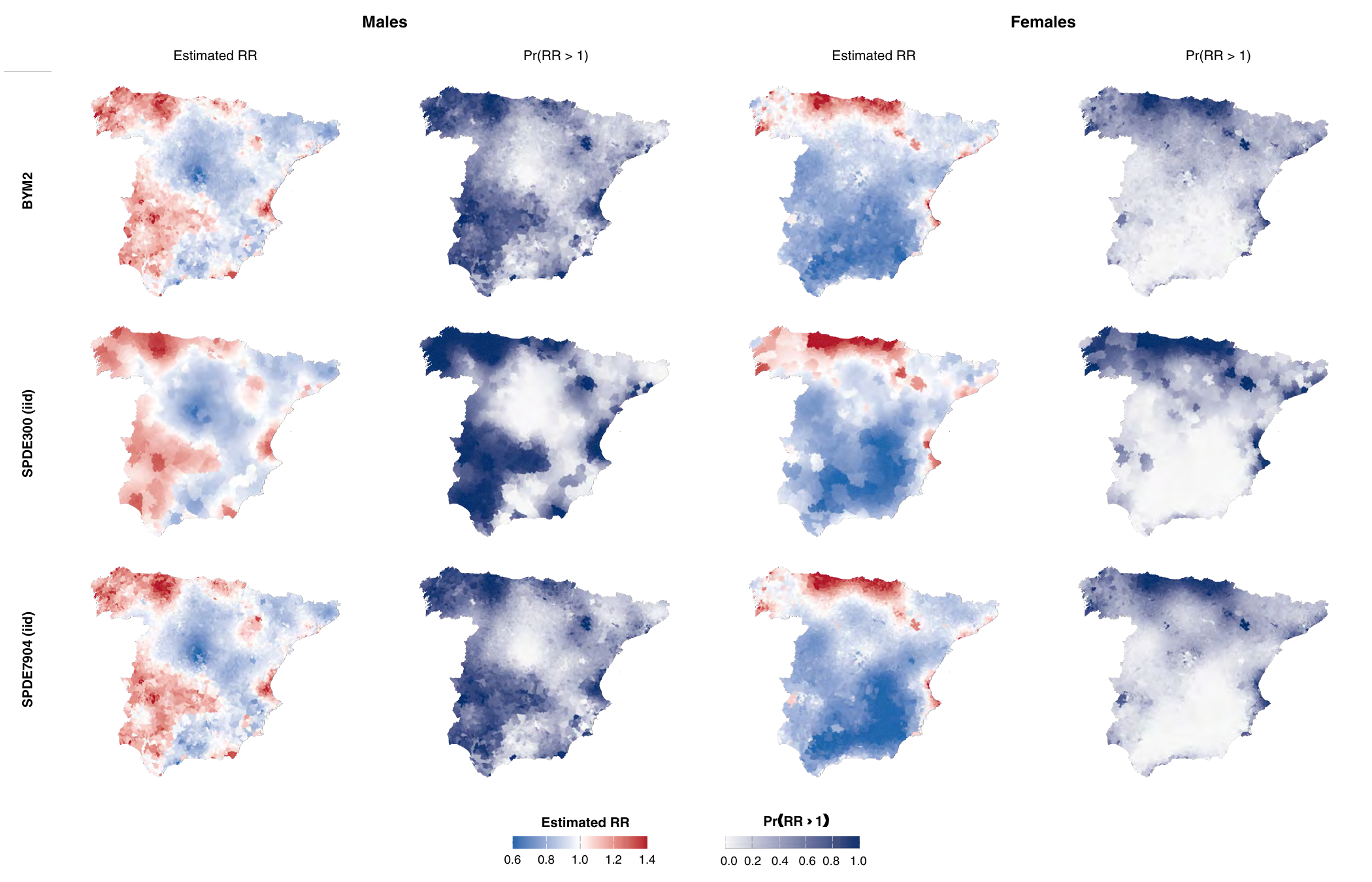}
	\caption{Estimated relative risk of lung cancer mortality and posterior probabilities in males and females aged 45--59.}
	\label{fig:supp_maps_45_59}
\end{figure}

\begin{figure}[H]
	\centering
	\includegraphics[
	width=\textwidth,
	height=0.80\textheight,
	keepaspectratio
	]{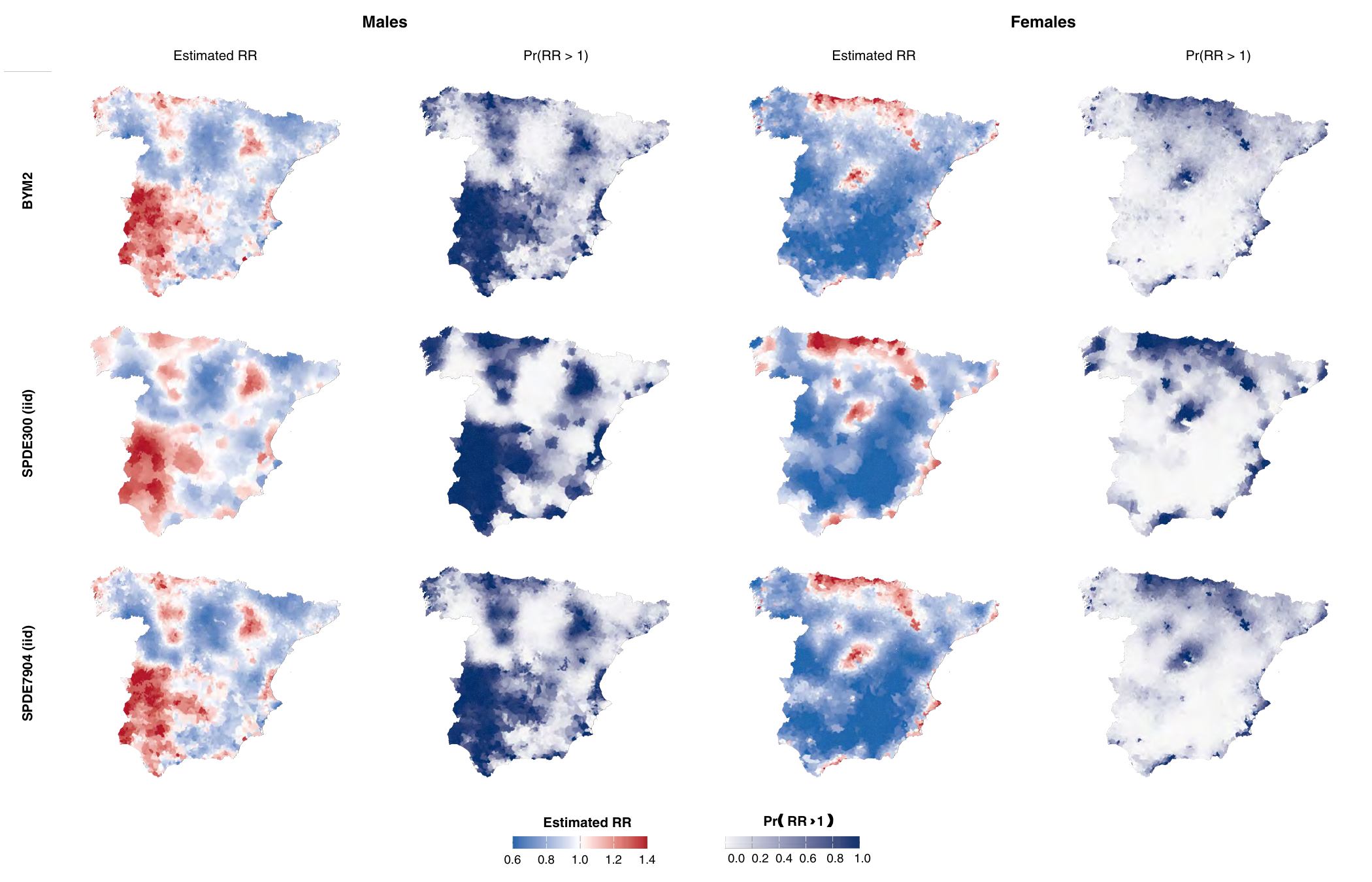}
	\caption{Estimated relative risk of lung cancer mortality and posterior probabilities in males and females aged 60--74.}
	\label{fig:supp_maps_60_74}
\end{figure}

Figures~\ref{fig:supp_maps_45_59} and~\ref{fig:supp_maps_60_74} show the estimated relative risks and posterior exceedance probabilities for males and females aged 45--59 and 60--74, respectively. Following the model comparison above, the SPDE maps correspond to the formulations including the iid component.

Among males aged 45--59, elevated risks are mainly observed in northern Spain, particularly in Galicia, Asturias, Cantabria and the Basque Country. Additional areas of elevated risk appear in western and southwestern Spain, including municipalities in Extremadura and western Andalusia. In the 60--74 age group, the western and southwestern pattern becomes more pronounced, particularly in Extremadura and in provinces of Huelva, Seville and Cádiz. Elevated risks also remain visible in parts of Galicia, Asturias and Cantabria. 

The spatial pattern among females is less extensive and more localized. For women aged 45--59, the clearest elevated risks are found along the northern coast, particularly in Galicia, Asturias, Cantabria and the Basque Country, with more isolated increases in Navarre and parts of Catalonia. Among women aged 60--74, elevated values remain visible in northern Spain, while additional areas appear around Madrid and along parts of the Valencian coast, especially in the province of Valencia.

The posterior probability maps support these general patterns. Among males, the strongest evidence of excess mortality is concentrated in northern Spain for ages 45--59 and becomes more extensive in western and southwestern Spain for ages 60--74. Among females, municipalities with high posterior probabilities are fewer and more geographically restricted, with the clearest concentrations in northern Spain and, in the 60--74 age group, around Madrid and the Valencian coast.

The persistence of the western and southwestern pattern among males may be consistent with the earlier and more widespread smoking epidemic in this population, as well as with the historical distribution of occupational and socioeconomic exposures. The more localized pattern among females may reflect the later diffusion of smoking among Spanish women, initially concentrated in more urbanized settings. However, the observed spatial patterns cannot be attributed exclusively to smoking, since occupational exposures, passive smoking, socioeconomic conditions and environmental factors may also contribute to local variation in mortality. 

\end{document}